\documentclass[twocolumn,superscriptaddress,longbibliography,aps,prx,
preprintnumbers]{revtex4-2}

\usepackage[normalem]{ulem}
\usepackage{graphicx}
\usepackage{bm}
\usepackage{color}
\usepackage{epstopdf}
\usepackage{amsmath}
\usepackage{amssymb}
\usepackage{epstopdf}
\usepackage{lipsum}

\usepackage{gensymb}

\usepackage{multirow}

\usepackage{scrextend}

\usepackage[urlcolor=blue,colorlinks=true,citecolor=blue,linkcolor=blue,pdfstartview={FitH},bookmarks=false]{hyperref}

\usepackage[dvipsnames]{xcolor}
\definecolor{mygreen}{rgb}{0.0, 0.6, 0.0}
\definecolor{pjorange}{rgb}{0.8, 0.3, 0.0}
\definecolor{jlblue}{rgb}{0.2, 0.5, 0.7}

\graphicspath{{fig/}{./fig/}{.}}

\begin{document}

\title{
Electronic and dynamical properties \\ of cobalt monogermanide CoGe phases under pressure
}

\author{Surajit Basak}
\email[e-mail: ]{surajit.basak@ifj.edu.pl}
\affiliation{\mbox{Institute of Nuclear Physics, Polish Academy of Sciences, W. E. Radzikowskiego 152, PL-31342 Krak\'{o}w, Poland}}

\author{Aksel~Kobia\l{}ka}
\affiliation{Department of Physics and Astronomy, Uppsala University, Uppsala SE-75120, Sweden}

\author{Ma\l{}gorzata~Sternik}
\affiliation{\mbox{Institute of Nuclear Physics, Polish Academy of Sciences, W. E. Radzikowskiego 152, PL-31342 Krak\'{o}w, Poland}}

\author{Jan~\L{}a\.{z}ewski}
\affiliation{\mbox{Institute of Nuclear Physics, Polish Academy of Sciences, W. E. Radzikowskiego 152, PL-31342 Krak\'{o}w, Poland}}

\author{Pawe\l{}~T.~Jochym}
\affiliation{\mbox{Institute of Nuclear Physics, Polish Academy of Sciences, W. E. Radzikowskiego 152, PL-31342 Krak\'{o}w, Poland}}

\author{Andrzej~M.~Ole\'{s}}
\affiliation{\mbox{Institute of Theoretical Physics, Jagiellonian University,
Prof. Stanis\l{}awa \L{}ojasiewicza 11, PL-30348 Krak\'{o}w, Poland}}

\author{Przemys\l{}aw~Piekarz}
\affiliation{\mbox{Institute of Nuclear Physics, Polish Academy of Sciences, W. E. Radzikowskiego 152, PL-31342 Krak\'{o}w, Poland}}

\author{Andrzej~Ptok}
\email[e-mail: ]{aptok@mmj.pl}
\affiliation{\mbox{Institute of Nuclear Physics, Polish Academy of Sciences, W. E. Radzikowskiego 152, PL-31342 Krak\'{o}w, Poland}}

\date{\today}

\begin{abstract}
We present the pressure dependence of the electronic and dynamical properties of six different CoGe phases: orthorhombic Cmmm, hexagonal P6/mmm and P$\bar{6}$2m, monoclinic C2/m, cubic P2$_{1}$3, and orthorhombic Pnma.
Using first-principles DFT calculations and the direct force-constants method, we study the dynamical stability of individual phases under external pressure.
We show that the orthorombic Cmmm and hexagonal P6/mmm structures are unstable over a broad pressure range and most pronounced imaginary phonon soft mode in both cases leads to a stable hexagonal P$\bar{6}$2m structure of the lowest ground-state energy of all studied phases at ambient and low (below $\sim 3$~GPa) external pressure.
Under these conditions, the cubic P2$_{1}$3 phase has the highest energy, however, together with monoclinic C2/m and orthorombic Pnma it is dynamically stable and all these three structures can potentially coexist as meta-stable phases.
Above $\sim 3$~GPa, the cubic P2$_{1}$3 phase becomes the most energetically favorable.
Fitting the Birch--Murnaghan equation of state we derive bulk modulus for all mentioned phases, which indicate relatively high resistance of CoGe to compression.
Such conclusions are confirmed by band structure calculations. Additionally, we show that electronic bands of the hexagonal P$\bar{6}$2m phase reveal characteristic features of the kagome-like structure, while in the cubic P2$_{1}$3 phase spectrum, one can locate spin-1 and double Weyl fermions.
In both cases, the external pressure induces the Lifshitz transition, related to the modification of the Fermi surface topology.
\end{abstract}

\maketitle

\section{Introduction}
\label{sec.intro}

Exploring the structure of materials is a fundamental first research step in physics, chemistry, and material science. 
The crystal structure as well as its other properties are inherently determined by bonds between atoms, molecules, or ions. 
However, a structure exposed to changing external conditions, such as temperature or pressure, may undergo a structural phase transition between different arrangements of atoms, causing the crystal symmetry change. 
Inspired by the recent discovery of several distinct structures of cobalt monogermanide, we 
performed an extensive study of stability and pressure dependence of different CoGe crystalline phases.

The CoGe binary compounds are an example of transition metal (TM) monogermanides~\cite{larchev.popova.82,larchev.popova.82,takizawa.sato.88,wilhelm.beanitz.11,grigoriev.potapova.13,ditusa.zhang.14,sidorov.petrova.18,zheng.rybakov.18,stolt.sigelko.18,grytsiuk.hoffmann.19,kamaeva.chtchelkatchev.21,beak.sidorov.22,chtchelkatchev.magnitskaya.20,bhan.schubert.60,richardson.67,morozkin.12}.
TM monogermanides are isostructural to TM monosilicides~\cite{jeong.pickett.04,grigoriev.chernyshov.09,grigoriev.chernyshov.10,pshenay.ivanov.18,rao.li.19,balasubramanian.manchanda.20}, but they are more difficult to grow and study~\cite{chtchelkatchev.magnitskaya.20}.
Nevertheless, such compounds have attracted a lot of attention because of the interesting properties observed in different phases.
Contrary to MnGe~\cite{ditusa.zhang.14,makarova.tsvyashchenko.12,kanazawa.kim.12} and similarly to FeGe, CoGe does not exhibit magnetic order~\cite{ditusa.zhang.14}, but can possess intrinsic spin Hall and spin Nernst effects~\cite{hsieh.prasad.22}.
The surface of CoGe is strongly active and shows bulk oxygen incorporation~\cite{pfau.trey.19}.
Similarly to RhGe~\cite{chtchelkatchev.magnitskaya.20} or FeGe~\cite{stolt.sigelko.18}, one can expect that the external conditions can cause a formation of various crystal structures.

Typically, the single crystal of CoGe is grown using a chemical vapor transport method~\cite{richardson.67}.
CoGe can also be synthesized in the cubic FeSi-type structure (B20) at high pressures and temperatures~\cite{larchev.popova.82,takizawa.sato.88}.
The cubic phase is a simple, low carrier density, metal, similar to CoSi~\cite{ditusa.zhang.14}.
Furthermore, the B20 phase was investigated by measuring the specific heat, resistivity, and $^{59}$Co nuclear magnetic resonance, which uncovered a phase transition at $13.7$~K~\cite{beak.sidorov.22}.
FeGe also crystallizes with the B20 structure and monoclinic phase.
At $893$~K, the cubic B20 phase transforms into the CoSn-type structure, which in turn undergoes a transition at $1013$~K to the high-temperature monoclinic polymorph, isostructural with CoGe~\cite{richardson.67}.
Furthermore, FeGe exhibits a hexagonal high-temperature P6/mmm structure, while decreasing temperature leads to the cubic P2$_{1}$3 structure~\cite{stolt.sigelko.18,kukolova.dimitrievska.21}.
High-pressure conditions should favor the cubic B20 structure which has the highest density of all monogermanides polymorphs~\cite{takizawa.sato.88,richardson.67}.
Nevertheless, the question of CoGe structure under pressure still remains open.
In this work we discuss the crystal stability of CoGe under pressure and its structural, electronic, and dynamical properties.

\begin{figure}[!ht]
\centering
\includegraphics[width=\columnwidth]{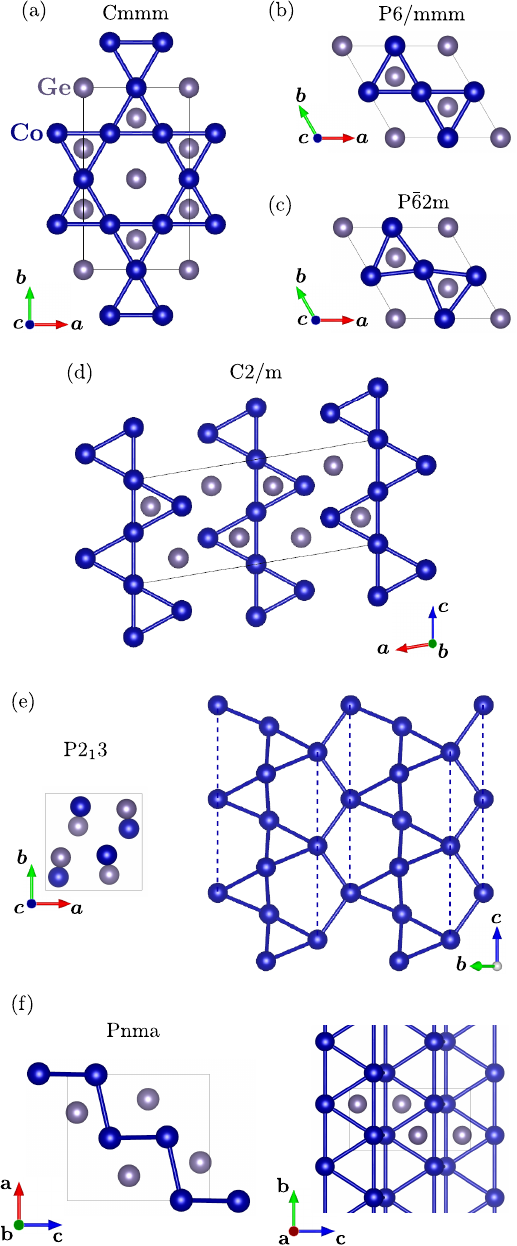}
\caption{
Overview of CoGe structures: 
orthorhombic Cmmm~(a),
hexagonal P6/mmm~(b) and P$\bar{6}$2m~(c),
monoclinic C2/m~(d),
cubic P2$_{1}$3~(e), and
orthorhombic Pnma~(f). 
\label{fig.schem}
}
\end{figure}

The paper is organized as follows.
Our results are presented and discussed in Sec.~\ref{sec.res}:
First, we analyze the investigated crystal structures (Sec.~\ref{sec.struc}).
Next, we describe lattice dynamics and system stability at zero pressure (Sec.~\ref{sec.stability}) and under external hydrostatic pressure (Sec.~\ref{sec.press}).
Afterwards, the electronic band structures of the most favorable crystal structures are presented (Sec.~\ref{sec.ele}).
Finally, we summarize and conclude our findings in Sec.~\ref{sec.sum}.
Details of the numerical calculation can be found in App.~\ref{sec.comp}


\section{System stability}
\label{sec.res}

\subsection{Crystal structures}
\label{sec.struc}

As we mentioned in the Introduction, TM monogermanides~\cite{larchev.popova.82,larchev.popova.82,takizawa.sato.88,wilhelm.beanitz.11,grigoriev.potapova.13,ditusa.zhang.14,sidorov.petrova.18,zheng.rybakov.18,stolt.sigelko.18,grytsiuk.hoffmann.19,kamaeva.chtchelkatchev.21,beak.sidorov.22} and TM monosilicides~\cite{jeong.pickett.04,grigoriev.chernyshov.09,grigoriev.chernyshov.10,pshenay.ivanov.18,rao.li.19,balasubramanian.manchanda.20} crystallize within several structures.
Early stage study of CoGe suggests existence of monoclinic C2/m~\cite{bhan.schubert.60,richardson.67,morozkin.12} and cubic P2$_{1}$3 (B20) structures~\cite{larchev.popova.82}.
Recently, theoretical studies have also predicted the hexagonal P$\bar{6}$2m structure~\cite{ptok.kobialka.21}.
Additionally, one should expect a strong impact of external conditions, such as temperature or pressure, on structure and stability of the material.
For example, the crystal structure of FeGe undergoes transformation from the hexagonal P6/mmm to cubic P2$_{1}$3 phase with decreasing temperature, around $625$~K~\cite{stolt.sigelko.18,kukolova.dimitrievska.21}.
Motivated by this, we take up here the topic of stability under pressure of a few plausible structures of CoGe.

Beyond the reported structures (C2/m, P2$_{1}$3, and P$\bar{6}$2m), we also examine following cases:
\begin{itemize}
\item Cmmm symmetry, which in {\sc OQMD} database~\cite{saal.kirklin.13,kirklin.saal.15} carries the lowest formation energy~\footnote{\url{https://oqmd.org/materials/composition/CoGe}};
\item P6/mmm symmetry -- hexagonal (B35) structure, well known from the FeGe system~\cite{richardson.67b}, but also reported for CoSn~\cite{larsson.haeberlein.96} and FeSn~\cite{waerenborgh.pereira.05};
\item Pnma symmetry --  MnP-like (B31) structure reported for $T$Ge (with $T=$ Ni, Pd, Ir, Pt)~\cite{pfisterer.schubert.50} and RhGe~\cite{geller.55}.
\end{itemize}
The last two symmetries were taken into account due to the chemical affinity of CoGe with other similar systems~\cite{wang.botti.21}.

All discussed structures of CoGe with different symmetries are presented in Fig.~\ref{fig.schem}.
In this group, we can find several similarities.
The orthorhombic Cmmm structure [Fig.~\ref{fig.schem}(a)], akin to the hexagonal P6/mmm and P$\bar{3}$2m symmetries [Fig.~\ref{fig.schem}(b) and~\ref{fig.schem}(c), respectively], contains two-dimensional (2D) kagome(-like) net of Co atoms.
The geometry of these structures implies the existence of exotic electronic dispersion relation, containing a flat band, characteristic for a 2D kagome lattice and reported e.g., for CoSn-like compounds~\cite{sales.yan.19,meier.du.20,kang.fang.20,kang.ye.20,lin.wang.20,huang.zheng.22,liu.li.20,han.inoue.21,sales.meier.21}.
The monoclinic C2/m and cubic P2$_{1}$3 structures [Figs~\ref{fig.schem}(d) and~\ref{fig.schem}(e), respectively] contain chains of apex-connected Co triangles.
A similar structure can also be found in the orthorhombic Pnma symmetry [Fig.~\ref{fig.schem}(f)] with a ladder formed of Co triangles.
In those cases, a quasi-one-dimensional (1D) structure should be reflected in the electronic band structure, in form of characteristic for 1D chains electronic bands.
Such feature was reported, e.g., in $A_{2}$Cr$_{3}$As$_{3}$ ~\mbox{\cite{wu.yang.15,wu.lee.15,jiang.cao.15,xu.wu.20}} or $A_{2}$Mo$_{3}$As$_{3}$~\cite{yang.feng.19,zhao.mu.20,lei.singh.21}, where $A=$K, Rb, and Cs.

\begin{figure*}
\centering
\includegraphics[width=\textwidth]{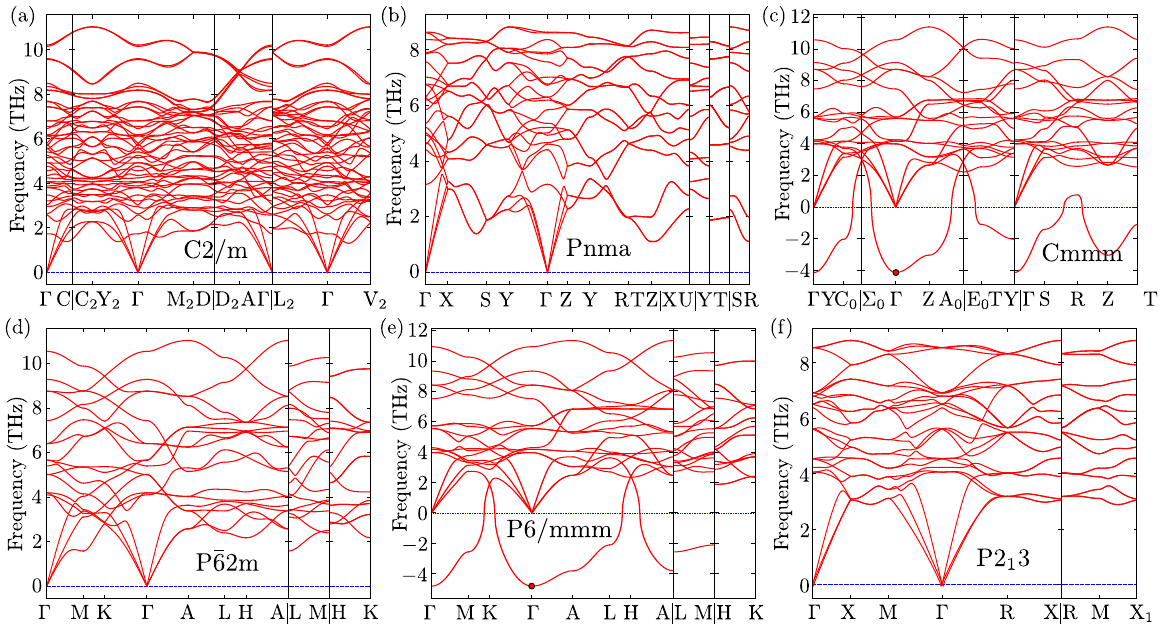}
\caption{
The phonon dispersion curves of CoGe along the high symmetry directions of considered structures at zero pressure. 
Corresponding symmetry groups are indicated in the graphs.
\label{fig.ph}
}
\end{figure*}

After optimization, we found the following lattice parameters:
\\[1ex]\textbf{The monoclinic C2/m} symmetry (space group No.~12): $a = 11.722$~\AA, $b = 3.763$~\AA, and $c = 4.930$~\AA\ ($\beta = 99.436^{\circ}$). 
The Wyckoff positions are: 2a (0,0,0), 4i (0.1964,0,0.3260), and 2c (0,0,1/2) for Co, and 4i (0.8169,0,0.1821) and 4i (0.5693,0,0.2807) for Ge.
The lattice constants of the relaxed structures are in good agreement with those reported experimentally for the monoclinic phase~\cite{takizawa.sato.88}: $a = 11.650$~\AA, $b = 3.807$~\AA, and $c = 4.945$~\AA\ ($\beta = 101.1^{\circ}$).
The lattice parameters are also comparable with the parameters reported for FeGe~\cite{richardson.67}: $a = 11.838$~\AA, $b = 3.937$~\AA, and $c = 4.934$~\AA\ ($\beta = 103.514^{\circ}$).
\\[1ex]\textbf{The orthorhombic Pnma} symmetry (space group No.~62):
$a = 5.410$~\AA, $b = 3.215$~\AA, and $c = 6.112$~\AA.
The Wyckoff positions are 4c ($-0.0045$,$1/4$,$0.2983$) for Co, and 4c ($0.8026$,$1/4$,$-0.0697$) for Ge.
\\[1ex]\textbf{The orthorhombic Cmmm} symmetry (space group No.~65): $a = 4.984$~\AA, $b = 8.632$~\AA, and $c = 3.884$~\AA.
The Wyckoff positions are 2b (1/2,0,0) and 4e (1/4,1/4,0) for Co, as well as 2a (0,0,0) and 4j (0,2/3,1/2) for Ge.
\\[1ex]\textbf{The hexagonal P$\bar{6}$2m} symmetry (space group No.~189): $a = b = 5.009$~\AA, and $c = 3.857$~\AA.
The Wyckoff positions are 3f ($0.4665$,$0$,$0$) for Co, and 1a (0,0,0) and 2d (1/3,2/3,0) for Ge.
\\[1ex]\textbf{The hexagonal P6/mmm} symmetry (space group No.~191): $a = b = 4.987$~\AA, and $c = 3.876$~\AA.
The Wyckoff positions are 3f (1/2,0,0) for Co, and 1a (0,0,0) and 2d (1/3,2/3,0) for Ge.
\\[1ex]\textbf{The cubic P2$_{1}$3} symmetry (space group No.~198): $a = b = c = 4.640$~\AA.
The Wyckoff positions are 4a~($0.6360$,$0.6360$,$0.6360$) for Co, and 4a ($0.3394$, $0.3394$, $0.3394$) for Ge.
The lattice constant is in a good agreement with the experimental one, $\sim 4.635$~\AA~\cite{takizawa.sato.88,ditusa.zhang.14,barman.mondal.20} and close to lattice constants of the similar monogermanides: $4.797$~\AA\ for MnGe~\cite{ditusa.zhang.14} and $4.70$~\AA\ for FeGe~\cite{richardson.67,tsvyashchenko.sidorov.16}.

\subsection{Zero pressure}
\label{sec.stability}

To check the system's stability, we calculate the phonon dispersion relations for the symmetries mentioned above (Fig.~\ref{fig.ph}).
Since the number of degrees of freedom of the primitive unit cell determines an amount of dispersion relations, the phonon spectrum of C2/m [Fig.~\ref{fig.ph}(a)] is the most complex.
Similar crystal structures [containing the kagome-like net, cf. Fig.~\ref{fig.schem}(a)--(c)] exhibit comparable phonon dispersion curves [cf. Fig.~\ref{fig.ph}(c)--(e)].
The phonon frequency ranges for all presented structures are analogous.

\begin{figure}[!t]
\centering
\includegraphics[width=\columnwidth]{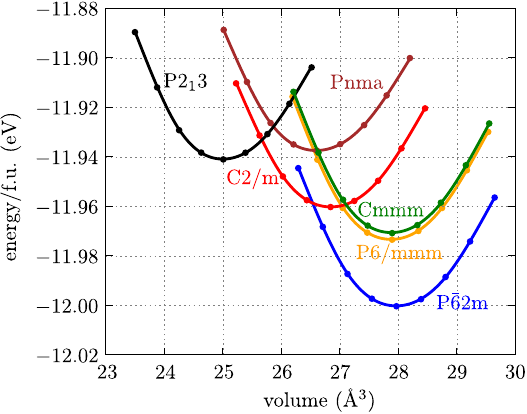}
\caption{
Volume dependence of the ground-state energy calculated at zero pressure for different structures of CoGe.
\label{fig.ene_vol}
}
\end{figure}

\begin{table}[!b]
\centering
\scriptsize
\caption{
\label{tab.param} 
The ground-state energy and equilibrium volume calculated per formula unit, as well as bulk modulus $B_{0}$ and its pressure derivative $B'_{0}$ fitted with the Birch--Murnaghan equation of state for different structures of CoGe. 
}
\begin{ruledtabular}
\begin{tabular}{ccccc}
Symmetry & $B_{0}$ & $B'_{0}$ & energy/f.u. & volume/f.u. \\
 &  (GPa) &  & (eV) & (\AA$^{3}$) \\
\hline
C2/m (SG:12) & $147.40$ & $3.90$ & $-11.966$ & $26.594$ \\
Pnma (SG:62) & $142.59$ & $5.76$ & $-11.934$ & $26.346$ \\
Cmmm (SG:65) & $158.62$ & $4.90$ & $-11.973$ & $27.655$ \\
P$\bar{6}$2m (SG:189) & $157.01$ & $4.82$ & $-11.999$ & $27.746$ \\
P6/mmm (SG:191) & $158.62$ & $4.82$ & $-11.971$ & $27.658$ \\
P2$_{1}$3 (SG:198) & $153.81$ & $6.65$ & $-11.941$ & $24.801$ \\
\end{tabular}
\end{ruledtabular}
\end{table}

Such similarities are also visible in the volume dependence of the ground state energy calculated for systems of different symmetries (Fig.~\ref{fig.ene_vol}).
All structures have a comparable volume and nearly the same (within 0.5\%) energy per one formula unit.
Fitting the Birch--Murnaghan equation of state~\cite{fu.ho.83}:
\begin{eqnarray}
\nonumber E(V) = E(V_{0}) + \frac{B_{0}V}{B'_{0}} \left( \frac{ (V_{0}/V)^{B'_{0}} }{ B'_{0}-1 } + 1 \right) - \frac{ V_{0} B_{0} }{ B'_{0} - 1 } , \\
\end{eqnarray}
to energy versus volume data, we found a bulk modulus $B_{0}$ and its pressure derivative $B'_{0}$ at the equilibrium volume $V_{0}$ (Tab.~\ref{tab.param}).
All symmetries are characterized by relatively large bulk modulus, which indicates a weak impact of external pressure on the system's mechanical properties.

Some of the structures discussed above can be eliminated at zero pressure due to instability of   
harmonic phonons.
This applies especially to Cmmm and P6/mmm structures, which show imaginary soft modes at the $\Gamma$ point.
Interestingly, in both structures, this soft mode is associated with the same deformation of the kagome-net, i.e.\ mutually opposite rotation of the Co triangles forming this sublattice [cf. Figs~\ref{fig.schem}(a) and~\ref{fig.schem}(b) with Fig.~\ref{fig.schem}(c)] around the $c$ axis~\cite{ptok.kobialka.21}.
After optimization, the distorted kagome lattice in P$\bar{6}$2m is stable [Fig.~\ref{fig.schem}(c)] without imaginary frequencies in the phonon spectrum [Fig.~\ref{fig.ph}(d)].
In the final structure, with the P$\bar{6}$2m symmetry, the Co-triangles within the kagome-like structure are rotated by $4^{\circ}$, which is close to the rotation angle observed experimentally in RhPb~\cite{ptok.meier.23}, i.e. $\sim 4.5^{\circ}$.

\begin{figure}[!b]
\centering
\includegraphics[width=\columnwidth]{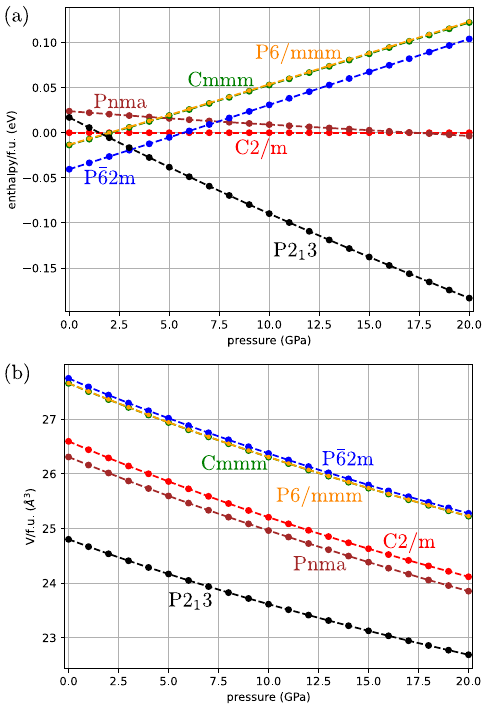}
\caption{
Pressure dependence of enthalpy (a) and volume (b) calculated (per formula unit) for different CoGe structures (as labeled).
\label{fig.press}
}
\end{figure}

The phonon dispersion curves of the cubic P2$_{1}$3 phase [Fig.~\ref{fig.ph}(f)] are similar to those reported for RhGe~\cite{chtchelkatchev.magnitskaya.20}.
Under zero pressure, the cubic P2$_{1}$3 (B20) phase has the highest energy among the reported structures (Fig.~\ref{fig.ene_vol}).
However, as mentioned earlier, the transition from the hexagonal (P6/mmm) to cubic (P2$_{1}$3) symmetry occurs in FeGe also due to temperature increase~\cite{stolt.sigelko.18,kukolova.dimitrievska.21}.

\begin{figure*}
\centering
\includegraphics[width=\textwidth]{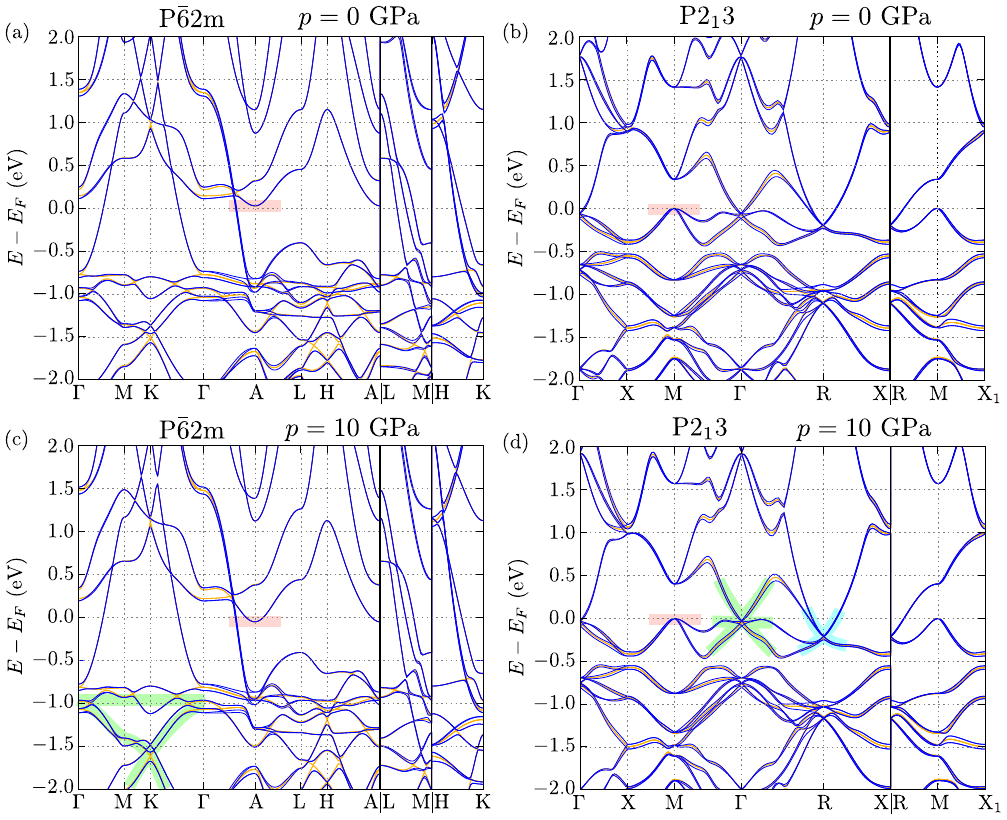}
\caption{
The top and bottom panels present the electronic band structure in the absence and presence of the external pressure for the hexagonal P$\bar{6}$2m and cubic P2$_{1}$3 phases (left and right column, respectively).
Orange and blue lines correspond to the results obtained with and without spin--orbit coupling, respectively.
The P$\bar{6}$2m band structure contains characteristic kagome-like features [marked by green background color on (c)], while the P2$_{1}$3 exhibits typical spin-1 and double Weyl fermions features [marked by green and blue underground colors on (d), respectively].
Increasing pressure (cf.\ top and bottom panels) causes the Lifshitz transition [places marked by red background color, at the A point in P$\bar{6}$2m and the M point in P2$_{1}$3 structure].
\label{fig.el_band}
}
\end{figure*}

\subsection{Role of external pressure}
\label{sec.press}

The stability of the system depends on external conditions.
Below, we discuss the effect of the external pressure on the system's stability.
The comparison of enthalpy (per formula unit) for the discussed symmetries is presented in Fig.~\ref{fig.press}(a).
As the reference level of energy, we choose the energy of the C2/m structure (red line).
At low pressures, the P$\bar{6}$2m structure is the most favorable energetically.
Then, above $\sim 3$~ GPa, the cubic phase has the lowest energy and should be preferred, which is in agreement with the previous predictions~\cite{takizawa.sato.88,richardson.67}.
However, regardless of the mutual ground-state energy relations, under specific conditions, crystal can grow in some meta-stable structures mentioned earlier (i.e.\ C2/m, Pnma, or P2$_{1}$3 structure).
Based on this, we expect that the experimentally reported monoclinic C2/m phase~\cite{richardson.67} can come from the cubic P2$_{1}$3 structure at low temperatures.

The unit cell of the unstable Cmmm phase can be constructed by doubling  P6/mmm or P$\bar{6}$2m unit cells.
For the Cmmm and P6/mmm symmetries, the ground-state energies and equilibrium volumes are mostly the same over the entire pressure range (cf.~green and orange lines in Figs~\ref{fig.ene_vol} and~\ref{fig.press}(a)).
However, the imaginary soft mode in the phonon spectra indicates existence of a structure with lower energy. 
Indeed, our group-theoretical analysis of both soft modes points out at the dynamically stable structure of P$\bar{6}$2m symmetry with energy systematically lower than those of the Cmmm and P6/mmm phases.
Even though all these structures have the same volume under pressure [cf.~green, orange, and blue lines on Fig.~\ref{fig.press}(b)] only the P$\bar{6}$2m structure is stable over the entire pressure range [cf.~blue line with green and orange lines on Fig.~\ref{fig.press}(a)].

As expected, due to the relatively large value of bulk modulus, volume (per formula unit) does not strongly depend on pressure [see Fig.~\ref{fig.press}(b)].
Therefore, independently of the structure symmetry (i.e.\ arrangement of atoms), the atomic density of systems is approximately the same and inversely proportional to volume.
Similarly, all structures exhibit similar compressibility, which is reflected in the relatively weak pressure dependence of volume (independently of the symmetry of the system).
For example, the hexagonal P$\bar{6}$2m structure under external pressure of $10$~GPa changes the lattice constants from $a = b = 5.009$~\AA\ and $c = 3.857$~\AA\ to $a = b = 4.914$~\AA\ and $c = 3.790$~\AA.
Similarly, the cubic P2$_{1}$3 phase lattice constant shrinks from $4.640$~\AA\ to $4.556$~\AA.
In both cases, the relative modification of the lattice constants induced by such pressure is around $\sim 2$~\%.

Due to small compressibility and volume modification, the orbital overlap does not change much.
Consequently, electronic band structures are unaffected by external pressure (see top and bottom panels in Fig.~\ref{fig.el_band}).
Also, there is only small variation in corresponding phonon dispersion relations (not shown).
For example, in the cubic P2$_{1}$3 structure only the frequency of the highest phonon mode at the $\Gamma$ point is apparently changing from $8.54$~THz to $9.29$~THz, while the main features of the other phonon dispersion curves remain unchanged.

From the above analysis we can conclude that, under some ``critical'' pressure (estimated from theoretical calculations as $\sim3$~GPa), the studied system transforms from the P$\bar{6}$2m to P2$_{1}$3 symmetry -- similarly to FeGe, which undergoes the phase transition from the hexagonal P6/mmm to cubic P2$_{1}$3 symmetry when temperature decreases~\cite{stolt.sigelko.18,kukolova.dimitrievska.21}.
During the described transition, the volume of CoGe increases by about $2.75$~\AA$^{3}$ per formula unit.

\begin{figure}[!b]
\centering
\includegraphics[width=0.8\columnwidth]{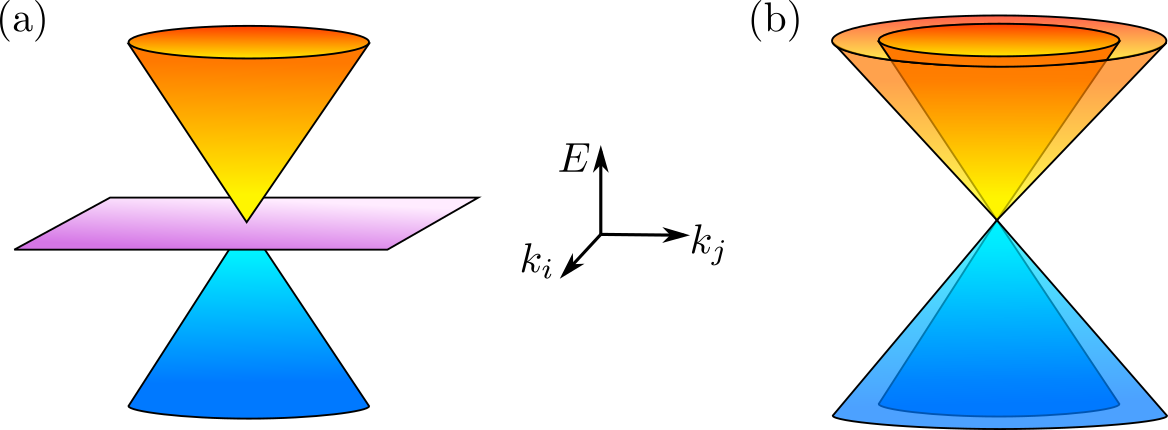}
\caption{
Schematic band dispersions in 3D $E$-$k$ space for the spin-1 fermion (a) and double Weyl fermion (b).
\label{fig.schemat}
}
\end{figure}

\section{Electronic properties}
\label{sec.ele}

For phases with the lowest energies on the enthalpy vs. pressure graph (Fig.~\ref{fig.press}), i.e. P$\bar{6}$2m and P2$_{1}$3, we investigate the electronic band structure.
Their dispersion relations (Fig.~\ref{fig.el_band}) are characteristic for these symmetries and, generally, a whole class of similar materials.
In the case of the hexagonal P$\bar{6}$2m phase, the electronic band structure exhibits the unique features of a system containing ``kagome'' net. 
For an ideal kagome lattice, the band structure contains a Dirac crossing at the K point, a strong van Hove singularity at the M point, and an almost flat band~\cite{sales.yan.19,meier.du.20,kang.fang.20,kang.ye.20,lin.wang.20,huang.zheng.22,liu.li.20,han.inoue.21,sales.meier.21}. 
We should notice that, contrary to the ideal 2D kagome lattice, where a perfectly flat band is realized, in the three-dimensional (3D) multi-orbital systems the kagome-related ``flat'' band has a finite bandwidth.
The nearly-flat band is mostly associated with the $d_{xz/yz}$ and $d_{xy/x^{2}-y^{2}}$ orbitals of Co atoms~\cite{kang.fang.20,meier.du.20,huang.zheng.22} forming the distorted kagome net [bands marked with the green background line in Fig.~\ref{fig.el_band}(c)].
Analogously to CoSn-like compounds, the flat bands are located around $-1$~eV~\cite{meier.du.20,sales.meier.21,ptok.meier.23}.
In case of the cubic P2$_{1}$3 phase, the electronic band structure exhibits characteristic features of TM monosilicide compounds~\cite{tang.zhou.17,chang.xu.17,sanchez.balopolski.19,pshenay.ivanov.18,takane.wang.19,schroter.pei.19,li.xu.19,rao.li.19,yao.manna.20,bose.narayan.21}, such as features of spin-1 fermions at the $\Gamma$ point, and double degenerate Weyl points at the R point [marked by green and cyan background lines in Fig.~\ref{fig.el_band}(d)].
The spin-1 fermions are related to the crossing of three doubly-degenerate bands (in the absence of spin--orbit coupling).
Similarly, one can observe the double Weyl point built by two Dirac-like cones centered at the same point.
Both band structures of spin-1 fermions and double Weyl point are presented schematically in Fig.~\ref{fig.schemat}.
As a consequence, a large Fermi arc is observed in the surface spectral function of CoGe with the cubic structure~\cite{barman.mondal.20}.

\begin{figure}[!t]
\centering
\includegraphics[width=\columnwidth]{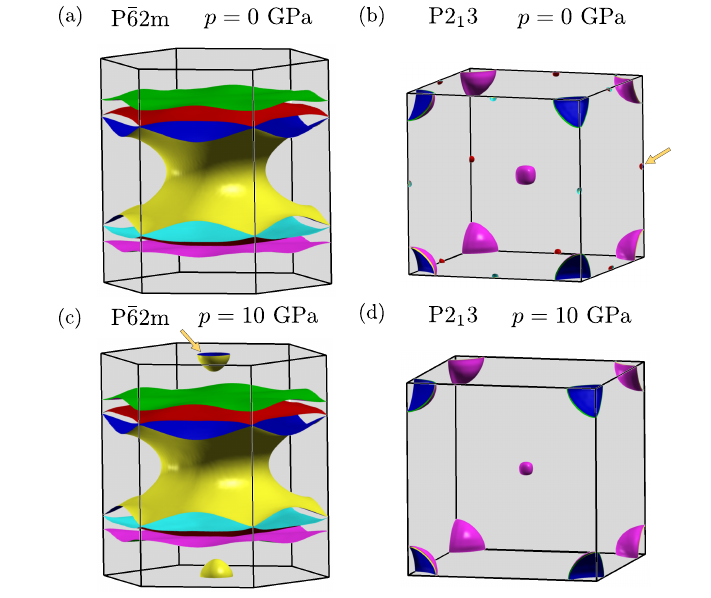}
\caption{
Modification of the Fermi surface by the external pressure for the hexagonal P$\bar{6}$2m and cubic P2$_{1}$3 phases (as labeled).
\label{fig.fs}
}
\end{figure}

The main features of the electronic band structure remain mostly unchanged under pressure.
Nevertheless, in both cases, external hydrostatic pressure leads to the Lifshitz transition~\cite{lifshitz.60}, i.e. change in the Fermi surface topology (see Fig.~\ref{fig.fs}).
In both structures, the compression shifts electronic bands and modifies the Fermi surface.
In the case of hexagonal P$\bar{6}$2m structure, new Fermi pockets emerge around the A point [cf. Fig~\ref{fig.fs}(a) and~\ref{fig.fs}(c)].
On the other hand, in the cubic P2$_{1}$3 structure the small Fermi pocket at the M point disappears under pressure [cf. Fig~\ref{fig.fs}(b) and~\ref{fig.fs}(d)].
Such modifications of the Fermi surface under pressure are related to a relatively small modification of the electronic band structure under pressure.
In the hexagonal structure, the bottom of the electron-like band at the A point is shifted to lower energies [cf. the electronic band structure at the A point, marked by a red background in Fig~\ref{fig.el_band}(a) and~\ref{fig.el_band}(c)].
Similarly, in the case of the cubic structure, the top of the hole-like band at the M point is also shifted to lower energies [cf. the electronic band structure at the M point, marked by the red background in Fig~\ref{fig.el_band}(b) and~\ref{fig.el_band}(d)].


\section{Summary}
\label{sec.sum}

In summary, we investigated the stability of several structures of cobalt monogermanide CoGe: monoclinic C2/m, orthorhombic Cmmm and Pnma, hexagonal P6/mmm and P$\bar{6}$2m, and cubic P2$_{1}$3.
From the study of lattice dynamics, we found that the monoclinic Cmmm and hexagonal P6/mmm structures are unstable and have the imaginary soft modes in the phonon spectra.
Based on group-theoretical analysis, we reveal that both soft modes lead to the same stable P$\bar{6}$2m structure, containing the distorted kagome lattice of Co atoms.
Surprisingly, under ambient pressure, the P$\bar{6}$2m structure has the lowest energy among studied phases.
The cubic P2$_{1}$3 structure is energetically favored under pressure above $\sim 3$~GPa.

We also discussed the electronic band structure of the most stable hexagonal P$\bar{6}$2m and cubic P2$_{1}$3 phases.
We demonstrated that the former one shows characteristic features of the compounds containing the kagome net, while the latter exhibits traits of the chiral cubic structure, such as spin-1 fermions and double Weyl fermions. 
In fact, the P$\bar{6}$2m structure contains the distorted kagome net of Co atoms, with two triangles forming the kagome-like net rotated in the opposite directions about $4^{\circ}$ around the $c$ axis.
Furthermore, we show that external pressure weakly affects the main features of the electronic band structure.
Nevertheless, the external hydrostatic pressure leads to the Lifshitz transition in both cases.



\begin{acknowledgments}
Some figures in this work were rendered using {\sc Vesta}~\cite{momma.izumi.11} and {\sc XCrySDen}~\cite{kokalj.99} software.
We kindly acknowledge support by National Science Centre (NCN, Poland) 
under Project No.~2021/43/B/ST3/02166. 
\end{acknowledgments}



\appendix

\section{Computational techniques}
\label{sec.comp}

The first-principles density functional theory (DFT) calculations were performed using the Vienna Ab initio Simulation Package ({\sc Vasp}) code~\cite{kresse.hafner.94,kresse.furthmuller.96,kresse.joubert.99} with the projector augmented-wave (PAW) potentials~\cite{blochl.94}.
For the exchange-correlation energy, the generalized gradient approximation (GGA) in the Perdew, Burke, and Ernzerhof for solids (PBEsol) parametrization was used~\cite{perdew.ruzsinszky.08}.
The energy cutoff for the plane-wave expansion was set to $350$~eV.

The optimization of the lattice constants and atom positions, including the spin--orbit coupling, was performed in the conventional unit cells.
As a convergence condition of the optimization loop, we took the energy change below $10^{-6}$~eV and $10^{-8}$~eV for the ionic and electronic degrees of freedom, respectively.
The following {\bf k}--point grids within the Monkhorst--Pack~\cite{monkhorst.pack.76} scheme were used for particular symmetries: $4 \times 12 \times 10$ for monoclinic C2/m, $6 \times 5 \times 3$ for orhoromic Pnma, $10 \times 6 \times 12$ for orhorombic Cmmm, $10 \times 10 \times 6$ for hexagonal P6/mmm and P$\bar{6}$2m, and $10 \times 10 \times 10$ for cubic P2$_{1}$3.
The symmetries of the system were analyzed using {\sc FindSym}~\cite{stokes.hatch.05} and {\sc Spglib}~\cite{togo.tanaka.18}, while momentum space analysis was performed with {\sc SeeK-path}~\cite{hinuma.pizzi.17}.

The dynamical properties were calculated using the direct {\it Parlinski--Li--Kawazoe} method~\cite{parlinski.li.97}, implemented in {\sc Phonopy} package~\cite{togo.chaput.23,togo.23}. 
Within this method, the interatomic force constants (IFC) are calculated from the Hellmann-Feynman (HF) forces generated by displacements of individual atoms inside the supercell.
In our calculations, we used the following supercell sizes for different symmetries:
$1 \times 3 \times 2$ for monoclinic C2/m, $2 \times 1 \times 3$ for orhoromic Pnma, $2 \times 3 \times 2$ for orhorombic Cmmm, $2 \times 2 \times 2$ for hexagonal P6/mmm and P$\bar{6}$2m, and $2 \times 2 \times 2$ for cubic P2$_{1}$3.
Phonon calculations were performed with the reduced $4 \times 4 \times 4$ {\bf k}-points grid.


\bibliography{biblio.bib}

\begin{thebibliography}{79}%
\makeatletter
\providecommand \@ifxundefined [1]{%
 \@ifx{#1\undefined}
}%
\providecommand \@ifnum [1]{%
 \ifnum #1\expandafter \@firstoftwo
 \else \expandafter \@secondoftwo
 \fi
}%
\providecommand \@ifx [1]{%
 \ifx #1\expandafter \@firstoftwo
 \else \expandafter \@secondoftwo
 \fi
}%
\providecommand \natexlab [1]{#1}%
\providecommand \enquote  [1]{``#1''}%
\providecommand \bibnamefont  [1]{#1}%
\providecommand \bibfnamefont [1]{#1}%
\providecommand \citenamefont [1]{#1}%
\providecommand \href@noop [0]{\@secondoftwo}%
\providecommand \href [0]{\begingroup \@sanitize@url \@href}%
\providecommand \@href[1]{\@@startlink{#1}\@@href}%
\providecommand \@@href[1]{\endgroup#1\@@endlink}%
\providecommand \@sanitize@url [0]{\catcode `\\12\catcode `\$12\catcode
  `\&12\catcode `\#12\catcode `\^12\catcode `\_12\catcode `\%12\relax}%
\providecommand \@@startlink[1]{}%
\providecommand \@@endlink[0]{}%
\providecommand \url  [0]{\begingroup\@sanitize@url \@url }%
\providecommand \@url [1]{\endgroup\@href {#1}{\urlprefix }}%
\providecommand \urlprefix  [0]{URL }%
\providecommand \Eprint [0]{\href }%
\providecommand \doibase [0]{https://doi.org/}%
\providecommand \selectlanguage [0]{\@gobble}%
\providecommand \bibinfo  [0]{\@secondoftwo}%
\providecommand \bibfield  [0]{\@secondoftwo}%
\providecommand \translation [1]{[#1]}%
\providecommand \BibitemOpen [0]{}%
\providecommand \bibitemStop [0]{}%
\providecommand \bibitemNoStop [0]{.\EOS\space}%
\providecommand \EOS [0]{\spacefactor3000\relax}%
\providecommand \BibitemShut  [1]{\csname bibitem#1\endcsname}%
\let\auto@bib@innerbib\@empty
\bibitem [{\citenamefont {Larchev}\ and\ \citenamefont
  {Popova}(1982)}]{larchev.popova.82}%
  \BibitemOpen
  \bibfield  {author} {\bibinfo {author} {\bibfnamefont {V.}~\bibnamefont
  {Larchev}}\ and\ \bibinfo {author} {\bibfnamefont {S.}~\bibnamefont
  {Popova}},\ }\bibfield  {title} {\bibinfo {title} {The polymorphism of
  transition metal monogermanides at high pressures and temperatures},\ }\href
  {https://doi.org/https://doi.org/10.1016/0022-5088(82)90040-6} {\bibfield
  {journal} {\bibinfo  {journal} {J. Less-Common Met.}\ }\textbf {\bibinfo
  {volume} {87}},\ \bibinfo {pages} {53} (\bibinfo {year} {1982})}\BibitemShut
  {NoStop}%
\bibitem [{\citenamefont {Takizawa}\ \emph {et~al.}(1988)\citenamefont
  {Takizawa}, \citenamefont {Sato}, \citenamefont {Endo},\ and\ \citenamefont
  {Shimada}}]{takizawa.sato.88}%
  \BibitemOpen
  \bibfield  {author} {\bibinfo {author} {\bibfnamefont {H.}~\bibnamefont
  {Takizawa}}, \bibinfo {author} {\bibfnamefont {T.}~\bibnamefont {Sato}},
  \bibinfo {author} {\bibfnamefont {T.}~\bibnamefont {Endo}},\ and\ \bibinfo
  {author} {\bibfnamefont {M.}~\bibnamefont {Shimada}},\ }\bibfield  {title}
  {\bibinfo {title} {High-pressure synthesis and electrical and magnetic
  properties of {MnGe} and {CoGe} with the cubic {B20} structure},\ }\href
  {https://doi.org/10.1016/0022-4596(88)90051-5} {\bibfield  {journal}
  {\bibinfo  {journal} {J. Solid State Chem.}\ }\textbf {\bibinfo {volume}
  {73}},\ \bibinfo {pages} {40} (\bibinfo {year} {1988})}\BibitemShut {NoStop}%
\bibitem [{\citenamefont {Wilhelm}\ \emph {et~al.}(2011)\citenamefont
  {Wilhelm}, \citenamefont {Baenitz}, \citenamefont {Schmidt}, \citenamefont
  {R\"o\ss{}ler}, \citenamefont {Leonov},\ and\ \citenamefont
  {Bogdanov}}]{wilhelm.beanitz.11}%
  \BibitemOpen
  \bibfield  {author} {\bibinfo {author} {\bibfnamefont {H.}~\bibnamefont
  {Wilhelm}}, \bibinfo {author} {\bibfnamefont {M.}~\bibnamefont {Baenitz}},
  \bibinfo {author} {\bibfnamefont {M.}~\bibnamefont {Schmidt}}, \bibinfo
  {author} {\bibfnamefont {U.~K.}\ \bibnamefont {R\"o\ss{}ler}}, \bibinfo
  {author} {\bibfnamefont {A.~A.}\ \bibnamefont {Leonov}},\ and\ \bibinfo
  {author} {\bibfnamefont {A.~N.}\ \bibnamefont {Bogdanov}},\ }\bibfield
  {title} {\bibinfo {title} {Precursor phenomena at the magnetic ordering of
  the cubic helimagnet {FeGe}},\ }\href
  {https://doi.org/10.1103/PhysRevLett.107.127203} {\bibfield  {journal}
  {\bibinfo  {journal} {Phys. Rev. Lett.}\ }\textbf {\bibinfo {volume} {107}},\
  \bibinfo {pages} {127203} (\bibinfo {year} {2011})}\BibitemShut {NoStop}%
\bibitem [{\citenamefont {Grigoriev}\ \emph {et~al.}(2013)\citenamefont
  {Grigoriev}, \citenamefont {Potapova}, \citenamefont {Siegfried},
  \citenamefont {Dyadkin}, \citenamefont {Moskvin}, \citenamefont {Dmitriev},
  \citenamefont {Menzel}, \citenamefont {Dewhurst}, \citenamefont {Chernyshov},
  \citenamefont {Sadykov}, \citenamefont {Fomicheva},\ and\ \citenamefont
  {Tsvyashchenko}}]{grigoriev.potapova.13}%
  \BibitemOpen
  \bibfield  {author} {\bibinfo {author} {\bibfnamefont {S.~V.}\ \bibnamefont
  {Grigoriev}}, \bibinfo {author} {\bibfnamefont {N.~M.}\ \bibnamefont
  {Potapova}}, \bibinfo {author} {\bibfnamefont {S.-A.}\ \bibnamefont
  {Siegfried}}, \bibinfo {author} {\bibfnamefont {V.~A.}\ \bibnamefont
  {Dyadkin}}, \bibinfo {author} {\bibfnamefont {E.~V.}\ \bibnamefont
  {Moskvin}}, \bibinfo {author} {\bibfnamefont {V.}~\bibnamefont {Dmitriev}},
  \bibinfo {author} {\bibfnamefont {D.}~\bibnamefont {Menzel}}, \bibinfo
  {author} {\bibfnamefont {C.~D.}\ \bibnamefont {Dewhurst}}, \bibinfo {author}
  {\bibfnamefont {D.}~\bibnamefont {Chernyshov}}, \bibinfo {author}
  {\bibfnamefont {R.~A.}\ \bibnamefont {Sadykov}}, \bibinfo {author}
  {\bibfnamefont {L.~N.}\ \bibnamefont {Fomicheva}},\ and\ \bibinfo {author}
  {\bibfnamefont {A.~V.}\ \bibnamefont {Tsvyashchenko}},\ }\bibfield  {title}
  {\bibinfo {title} {Chiral properties of structure and magnetism in
  {Mn$_{1-x}$Fe$_{x}$Ge} compounds: When the left and the right are fighting,
  who wins?},\ }\href {https://doi.org/10.1103/PhysRevLett.110.207201}
  {\bibfield  {journal} {\bibinfo  {journal} {Phys. Rev. Lett.}\ }\textbf
  {\bibinfo {volume} {110}},\ \bibinfo {pages} {207201} (\bibinfo {year}
  {2013})}\BibitemShut {NoStop}%
\bibitem [{\citenamefont {DiTusa}\ \emph {et~al.}(2014)\citenamefont {DiTusa},
  \citenamefont {Zhang}, \citenamefont {Yamaura}, \citenamefont {Xiong},
  \citenamefont {Prestigiacomo}, \citenamefont {Fulfer}, \citenamefont {Adams},
  \citenamefont {Brickson}, \citenamefont {Browne}, \citenamefont {Capan},
  \citenamefont {Fisk},\ and\ \citenamefont {Chan}}]{ditusa.zhang.14}%
  \BibitemOpen
  \bibfield  {author} {\bibinfo {author} {\bibfnamefont {J.~F.}\ \bibnamefont
  {DiTusa}}, \bibinfo {author} {\bibfnamefont {S.~B.}\ \bibnamefont {Zhang}},
  \bibinfo {author} {\bibfnamefont {K.}~\bibnamefont {Yamaura}}, \bibinfo
  {author} {\bibfnamefont {Y.}~\bibnamefont {Xiong}}, \bibinfo {author}
  {\bibfnamefont {J.~C.}\ \bibnamefont {Prestigiacomo}}, \bibinfo {author}
  {\bibfnamefont {B.~W.}\ \bibnamefont {Fulfer}}, \bibinfo {author}
  {\bibfnamefont {P.~W.}\ \bibnamefont {Adams}}, \bibinfo {author}
  {\bibfnamefont {M.~I.}\ \bibnamefont {Brickson}}, \bibinfo {author}
  {\bibfnamefont {D.~A.}\ \bibnamefont {Browne}}, \bibinfo {author}
  {\bibfnamefont {C.}~\bibnamefont {Capan}}, \bibinfo {author} {\bibfnamefont
  {Z.}~\bibnamefont {Fisk}},\ and\ \bibinfo {author} {\bibfnamefont {J.~Y.}\
  \bibnamefont {Chan}},\ }\bibfield  {title} {\bibinfo {title} {Magnetic,
  thermodynamic, and electrical transport properties of the noncentrosymmetric
  {B20} germanides {MnGe} and {CoGe}},\ }\href
  {https://doi.org/10.1103/PhysRevB.90.144404} {\bibfield  {journal} {\bibinfo
  {journal} {Phys. Rev. B}\ }\textbf {\bibinfo {volume} {90}},\ \bibinfo
  {pages} {144404} (\bibinfo {year} {2014})}\BibitemShut {NoStop}%
\bibitem [{\citenamefont {Sidorov}\ \emph {et~al.}(2018)\citenamefont
  {Sidorov}, \citenamefont {Petrova}, \citenamefont {Chtchelkatchev},
  \citenamefont {Magnitskaya}, \citenamefont {Fomicheva}, \citenamefont
  {Salamatin}, \citenamefont {Nikolaev}, \citenamefont {Zibrov}, \citenamefont
  {Wilhelm}, \citenamefont {Rogalev},\ and\ \citenamefont
  {Tsvyashchenko}}]{sidorov.petrova.18}%
  \BibitemOpen
  \bibfield  {author} {\bibinfo {author} {\bibfnamefont {V.~A.}\ \bibnamefont
  {Sidorov}}, \bibinfo {author} {\bibfnamefont {A.~E.}\ \bibnamefont
  {Petrova}}, \bibinfo {author} {\bibfnamefont {N.~M.}\ \bibnamefont
  {Chtchelkatchev}}, \bibinfo {author} {\bibfnamefont {M.~V.}\ \bibnamefont
  {Magnitskaya}}, \bibinfo {author} {\bibfnamefont {L.~N.}\ \bibnamefont
  {Fomicheva}}, \bibinfo {author} {\bibfnamefont {D.~A.}\ \bibnamefont
  {Salamatin}}, \bibinfo {author} {\bibfnamefont {A.~V.}\ \bibnamefont
  {Nikolaev}}, \bibinfo {author} {\bibfnamefont {I.~P.}\ \bibnamefont
  {Zibrov}}, \bibinfo {author} {\bibfnamefont {F.}~\bibnamefont {Wilhelm}},
  \bibinfo {author} {\bibfnamefont {A.}~\bibnamefont {Rogalev}},\ and\ \bibinfo
  {author} {\bibfnamefont {A.~V.}\ \bibnamefont {Tsvyashchenko}},\ }\bibfield
  {title} {\bibinfo {title} {Magnetic, electronic, and transport properties of
  the high-pressure-synthesized chiral magnets {Mn$_{1-x}$Rh$_{x}$Ge}},\ }\href
  {https://doi.org/10.1103/PhysRevB.98.125121} {\bibfield  {journal} {\bibinfo
  {journal} {Phys. Rev. B}\ }\textbf {\bibinfo {volume} {98}},\ \bibinfo
  {pages} {125121} (\bibinfo {year} {2018})}\BibitemShut {NoStop}%
\bibitem [{\citenamefont {Zheng}\ \emph {et~al.}(2018)\citenamefont {Zheng},
  \citenamefont {Rybakov}, \citenamefont {Borisov}, \citenamefont {Song},
  \citenamefont {Wang}, \citenamefont {Li}, \citenamefont {Du}, \citenamefont
  {Kiselev}, \citenamefont {Caron}, \citenamefont {Kov{\'a}cs}, \citenamefont
  {Tian}, \citenamefont {Zhang}, \citenamefont {Bl{\"u}gel},\ and\
  \citenamefont {Dunin-Borkowski}}]{zheng.rybakov.18}%
  \BibitemOpen
  \bibfield  {author} {\bibinfo {author} {\bibfnamefont {F.}~\bibnamefont
  {Zheng}}, \bibinfo {author} {\bibfnamefont {F.~N.}\ \bibnamefont {Rybakov}},
  \bibinfo {author} {\bibfnamefont {A.~B.}\ \bibnamefont {Borisov}}, \bibinfo
  {author} {\bibfnamefont {D.}~\bibnamefont {Song}}, \bibinfo {author}
  {\bibfnamefont {S.}~\bibnamefont {Wang}}, \bibinfo {author} {\bibfnamefont
  {Z.-A.}\ \bibnamefont {Li}}, \bibinfo {author} {\bibfnamefont
  {H.}~\bibnamefont {Du}}, \bibinfo {author} {\bibfnamefont {N.~S.}\
  \bibnamefont {Kiselev}}, \bibinfo {author} {\bibfnamefont {J.}~\bibnamefont
  {Caron}}, \bibinfo {author} {\bibfnamefont {A.}~\bibnamefont {Kov{\'a}cs}},
  \bibinfo {author} {\bibfnamefont {M.}~\bibnamefont {Tian}}, \bibinfo {author}
  {\bibfnamefont {Y.}~\bibnamefont {Zhang}}, \bibinfo {author} {\bibfnamefont
  {S.}~\bibnamefont {Bl{\"u}gel}},\ and\ \bibinfo {author} {\bibfnamefont
  {R.~E.}\ \bibnamefont {Dunin-Borkowski}},\ }\bibfield  {title} {\bibinfo
  {title} {Experimental observation of chiral magnetic bobbers in {B20}-type
  {FeGe}},\ }\href {https://doi.org/10.1038/s41565-018-0093-3} {\bibfield
  {journal} {\bibinfo  {journal} {Nature Nanotech.}\ }\textbf {\bibinfo
  {volume} {13}},\ \bibinfo {pages} {451} (\bibinfo {year} {2018})}\BibitemShut
  {NoStop}%
\bibitem [{\citenamefont {Stolt}\ \emph {et~al.}(2018)\citenamefont {Stolt},
  \citenamefont {Sigelko}, \citenamefont {Mathur},\ and\ \citenamefont
  {Jin}}]{stolt.sigelko.18}%
  \BibitemOpen
  \bibfield  {author} {\bibinfo {author} {\bibfnamefont {M.~J.}\ \bibnamefont
  {Stolt}}, \bibinfo {author} {\bibfnamefont {X.}~\bibnamefont {Sigelko}},
  \bibinfo {author} {\bibfnamefont {N.}~\bibnamefont {Mathur}},\ and\ \bibinfo
  {author} {\bibfnamefont {S.}~\bibnamefont {Jin}},\ }\bibfield  {title}
  {\bibinfo {title} {Chemical pressure stabilization of the cubic {B20}
  structure in skyrmion hosting {Fe$_{1-x}$Co$_{x}$Ge} alloys},\ }\href
  {https://doi.org/10.1021/acs.chemmater.7b05261} {\bibfield  {journal}
  {\bibinfo  {journal} {Chem. Mater.}\ }\textbf {\bibinfo {volume} {30}},\
  \bibinfo {pages} {1146} (\bibinfo {year} {2018})}\BibitemShut {NoStop}%
\bibitem [{\citenamefont {Grytsiuk}\ \emph {et~al.}(2019)\citenamefont
  {Grytsiuk}, \citenamefont {Hoffmann}, \citenamefont {Hanke}, \citenamefont
  {Mavropoulos}, \citenamefont {Mokrousov}, \citenamefont {Bihlmayer},\ and\
  \citenamefont {Bl\"ugel}}]{grytsiuk.hoffmann.19}%
  \BibitemOpen
  \bibfield  {author} {\bibinfo {author} {\bibfnamefont {S.}~\bibnamefont
  {Grytsiuk}}, \bibinfo {author} {\bibfnamefont {M.}~\bibnamefont {Hoffmann}},
  \bibinfo {author} {\bibfnamefont {J.-P.}\ \bibnamefont {Hanke}}, \bibinfo
  {author} {\bibfnamefont {P.}~\bibnamefont {Mavropoulos}}, \bibinfo {author}
  {\bibfnamefont {Y.}~\bibnamefont {Mokrousov}}, \bibinfo {author}
  {\bibfnamefont {G.}~\bibnamefont {Bihlmayer}},\ and\ \bibinfo {author}
  {\bibfnamefont {S.}~\bibnamefont {Bl\"ugel}},\ }\bibfield  {title} {\bibinfo
  {title} {Ab initio analysis of magnetic properties of the prototype {B20}
  chiral magnet {FeGe}},\ }\href {https://doi.org/10.1103/PhysRevB.100.214406}
  {\bibfield  {journal} {\bibinfo  {journal} {Phys. Rev. B}\ }\textbf {\bibinfo
  {volume} {100}},\ \bibinfo {pages} {214406} (\bibinfo {year}
  {2019})}\BibitemShut {NoStop}%
\bibitem [{\citenamefont {Kamaeva}\ \emph {et~al.}(2021)\citenamefont
  {Kamaeva}, \citenamefont {Chtchelkatchev}, \citenamefont {Suslov},
  \citenamefont {Magnitskaya},\ and\ \citenamefont
  {Tsvyashchenko}}]{kamaeva.chtchelkatchev.21}%
  \BibitemOpen
  \bibfield  {author} {\bibinfo {author} {\bibfnamefont {L.~V.}\ \bibnamefont
  {Kamaeva}}, \bibinfo {author} {\bibfnamefont {N.~M.}\ \bibnamefont
  {Chtchelkatchev}}, \bibinfo {author} {\bibfnamefont {A.~A.}\ \bibnamefont
  {Suslov}}, \bibinfo {author} {\bibfnamefont {M.~V.}\ \bibnamefont
  {Magnitskaya}},\ and\ \bibinfo {author} {\bibfnamefont {A.~V.}\ \bibnamefont
  {Tsvyashchenko}},\ }\bibfield  {title} {\bibinfo {title} {Structural and
  thermal stability of {B20}-type high-pressure phases {FeGe} and {MnGe}},\
  }\href {https://doi.org/10.1016/j.jallcom.2021.161565} {\bibfield  {journal}
  {\bibinfo  {journal} {J. Alloys Compd.}\ }\textbf {\bibinfo {volume} {888}},\
  \bibinfo {pages} {161565} (\bibinfo {year} {2021})}\BibitemShut {NoStop}%
\bibitem [{\citenamefont {Baek}\ \emph {et~al.}(2022)\citenamefont {Baek},
  \citenamefont {Sidorov}, \citenamefont {Nikolaev}, \citenamefont {Klimczuk},
  \citenamefont {Ronning},\ and\ \citenamefont
  {Tsvyashchenko}}]{beak.sidorov.22}%
  \BibitemOpen
  \bibfield  {author} {\bibinfo {author} {\bibfnamefont {S.-H.}\ \bibnamefont
  {Baek}}, \bibinfo {author} {\bibfnamefont {V.~A.}\ \bibnamefont {Sidorov}},
  \bibinfo {author} {\bibfnamefont {A.~V.}\ \bibnamefont {Nikolaev}}, \bibinfo
  {author} {\bibfnamefont {T.}~\bibnamefont {Klimczuk}}, \bibinfo {author}
  {\bibfnamefont {F.}~\bibnamefont {Ronning}},\ and\ \bibinfo {author}
  {\bibfnamefont {A.~V.}\ \bibnamefont {Tsvyashchenko}},\ }\bibfield  {title}
  {\bibinfo {title} {Possible quadrupole-order-driven
  commensurate-incommensurate phase transition in {B20} {CoGe}},\ }\href
  {https://doi.org/10.1103/PhysRevB.105.165132} {\bibfield  {journal} {\bibinfo
   {journal} {Phys. Rev. B}\ }\textbf {\bibinfo {volume} {105}},\ \bibinfo
  {pages} {165132} (\bibinfo {year} {2022})}\BibitemShut {NoStop}%
\bibitem [{\citenamefont {Chtchelkatchev}\ \emph {et~al.}(2020)\citenamefont
  {Chtchelkatchev}, \citenamefont {Magnitskaya},\ and\ \citenamefont
  {Tsvyashchenko}}]{chtchelkatchev.magnitskaya.20}%
  \BibitemOpen
  \bibfield  {author} {\bibinfo {author} {\bibfnamefont {N.~M.}\ \bibnamefont
  {Chtchelkatchev}}, \bibinfo {author} {\bibfnamefont {M.~V.}\ \bibnamefont
  {Magnitskaya}},\ and\ \bibinfo {author} {\bibfnamefont {A.~V.}\ \bibnamefont
  {Tsvyashchenko}},\ }\bibfield  {title} {\bibinfo {title} {Ab initio study of
  noncentrosymmetric transition-metal monogermanide {B20-RhGe} synthesized at
  high temperature and pressure},\ }\href
  {https://doi.org/10.1140/epjst/e2019-900114-y} {\bibfield  {journal}
  {\bibinfo  {journal} {Eur. Phys. J. Spec. Top}\ }\textbf {\bibinfo {volume}
  {229}},\ \bibinfo {pages} {167} (\bibinfo {year} {2020})}\BibitemShut
  {NoStop}%
\bibitem [{\citenamefont {Bhan}\ and\ \citenamefont
  {Schubert}(1960)}]{bhan.schubert.60}%
  \BibitemOpen
  \bibfield  {author} {\bibinfo {author} {\bibfnamefont {S.}~\bibnamefont
  {Bhan}}\ and\ \bibinfo {author} {\bibfnamefont {K.}~\bibnamefont
  {Schubert}},\ }\bibfield  {title} {\bibinfo {title} {Zum aufbau der systeme
  kobalt-germanium, rhodium-silizium sowie einiger verwandter legierungen},\
  }\href {https://doi.org/doi:10.1515/ijmr-1960-510604} {\bibfield  {journal}
  {\bibinfo  {journal} {International Journal of Materials Research}\ }\textbf
  {\bibinfo {volume} {51}},\ \bibinfo {pages} {327} (\bibinfo {year}
  {1960})}\BibitemShut {NoStop}%
\bibitem [{\citenamefont {Richardson}(1967{\natexlab{a}})}]{richardson.67}%
  \BibitemOpen
  \bibfield  {author} {\bibinfo {author} {\bibfnamefont {M.~W.}\ \bibnamefont
  {Richardson}},\ }\bibfield  {title} {\bibinfo {title} {Crystal structure
  refinements of the {B 20} and monoclinic ({CoGe}-type) polymorphs of
  {FeGe}},\ }\href {https://doi.org/10.3891/acta.chem.scand.21-0753} {\bibfield
   {journal} {\bibinfo  {journal} {Acta Chem. Scand.}\ }\textbf {\bibinfo
  {volume} {21}},\ \bibinfo {pages} {753} (\bibinfo {year}
  {1967}{\natexlab{a}})}\BibitemShut {NoStop}%
\bibitem [{\citenamefont {Morozkin}(2012)}]{morozkin.12}%
  \BibitemOpen
  \bibfield  {author} {\bibinfo {author} {\bibfnamefont {A.}~\bibnamefont
  {Morozkin}},\ }\bibfield  {title} {\bibinfo {title} {{Gd--Co--Ge} system at
  870/1070 {K}},\ }\href {https://doi.org/10.1016/j.intermet.2012.03.001}
  {\bibfield  {journal} {\bibinfo  {journal} {Intermetallics}\ }\textbf
  {\bibinfo {volume} {25}},\ \bibinfo {pages} {136} (\bibinfo {year}
  {2012})}\BibitemShut {NoStop}%
\bibitem [{\citenamefont {Jeong}\ and\ \citenamefont
  {Pickett}(2004)}]{jeong.pickett.04}%
  \BibitemOpen
  \bibfield  {author} {\bibinfo {author} {\bibfnamefont {T.}~\bibnamefont
  {Jeong}}\ and\ \bibinfo {author} {\bibfnamefont {W.~E.}\ \bibnamefont
  {Pickett}},\ }\bibfield  {title} {\bibinfo {title} {Implications of the {B20}
  crystal structure for the magnetoelectronic structure of {MnSi}},\ }\href
  {https://doi.org/10.1103/PhysRevB.70.075114} {\bibfield  {journal} {\bibinfo
  {journal} {Phys. Rev. B}\ }\textbf {\bibinfo {volume} {70}},\ \bibinfo
  {pages} {075114} (\bibinfo {year} {2004})}\BibitemShut {NoStop}%
\bibitem [{\citenamefont {Grigoriev}\ \emph {et~al.}(2009)\citenamefont
  {Grigoriev}, \citenamefont {Chernyshov}, \citenamefont {Dyadkin},
  \citenamefont {Dmitriev}, \citenamefont {Maleyev}, \citenamefont {Moskvin},
  \citenamefont {Menzel}, \citenamefont {Schoenes},\ and\ \citenamefont
  {Eckerlebe}}]{grigoriev.chernyshov.09}%
  \BibitemOpen
  \bibfield  {author} {\bibinfo {author} {\bibfnamefont {S.~V.}\ \bibnamefont
  {Grigoriev}}, \bibinfo {author} {\bibfnamefont {D.}~\bibnamefont
  {Chernyshov}}, \bibinfo {author} {\bibfnamefont {V.~A.}\ \bibnamefont
  {Dyadkin}}, \bibinfo {author} {\bibfnamefont {V.}~\bibnamefont {Dmitriev}},
  \bibinfo {author} {\bibfnamefont {S.~V.}\ \bibnamefont {Maleyev}}, \bibinfo
  {author} {\bibfnamefont {E.~V.}\ \bibnamefont {Moskvin}}, \bibinfo {author}
  {\bibfnamefont {D.}~\bibnamefont {Menzel}}, \bibinfo {author} {\bibfnamefont
  {J.}~\bibnamefont {Schoenes}},\ and\ \bibinfo {author} {\bibfnamefont
  {H.}~\bibnamefont {Eckerlebe}},\ }\bibfield  {title} {\bibinfo {title}
  {Crystal handedness and spin helix chirality in {Fe$_{1-x}$Co$_{x}$Si}},\
  }\href {https://doi.org/10.1103/PhysRevLett.102.037204} {\bibfield  {journal}
  {\bibinfo  {journal} {Phys. Rev. Lett.}\ }\textbf {\bibinfo {volume} {102}},\
  \bibinfo {pages} {037204} (\bibinfo {year} {2009})}\BibitemShut {NoStop}%
\bibitem [{\citenamefont {Grigoriev}\ \emph {et~al.}(2010)\citenamefont
  {Grigoriev}, \citenamefont {Chernyshov}, \citenamefont {Dyadkin},
  \citenamefont {Dmitriev}, \citenamefont {Moskvin}, \citenamefont {Lamago},
  \citenamefont {Wolf}, \citenamefont {Menzel}, \citenamefont {Schoenes},
  \citenamefont {Maleyev},\ and\ \citenamefont
  {Eckerlebe}}]{grigoriev.chernyshov.10}%
  \BibitemOpen
  \bibfield  {author} {\bibinfo {author} {\bibfnamefont {S.~V.}\ \bibnamefont
  {Grigoriev}}, \bibinfo {author} {\bibfnamefont {D.}~\bibnamefont
  {Chernyshov}}, \bibinfo {author} {\bibfnamefont {V.~A.}\ \bibnamefont
  {Dyadkin}}, \bibinfo {author} {\bibfnamefont {V.}~\bibnamefont {Dmitriev}},
  \bibinfo {author} {\bibfnamefont {E.~V.}\ \bibnamefont {Moskvin}}, \bibinfo
  {author} {\bibfnamefont {D.}~\bibnamefont {Lamago}}, \bibinfo {author}
  {\bibfnamefont {T.}~\bibnamefont {Wolf}}, \bibinfo {author} {\bibfnamefont
  {D.}~\bibnamefont {Menzel}}, \bibinfo {author} {\bibfnamefont
  {J.}~\bibnamefont {Schoenes}}, \bibinfo {author} {\bibfnamefont {S.~V.}\
  \bibnamefont {Maleyev}},\ and\ \bibinfo {author} {\bibfnamefont
  {H.}~\bibnamefont {Eckerlebe}},\ }\bibfield  {title} {\bibinfo {title}
  {Interplay between crystalline chirality and magnetic structure in
  {Mn$_{1-x}$Fe$_{x}$Si}},\ }\href {https://doi.org/10.1103/PhysRevB.81.012408}
  {\bibfield  {journal} {\bibinfo  {journal} {Phys. Rev. B}\ }\textbf {\bibinfo
  {volume} {81}},\ \bibinfo {pages} {012408} (\bibinfo {year}
  {2010})}\BibitemShut {NoStop}%
\bibitem [{\citenamefont {Pshenay-Severin}\ \emph {et~al.}(2018)\citenamefont
  {Pshenay-Severin}, \citenamefont {Ivanov}, \citenamefont {Burkov},\ and\
  \citenamefont {Burkov}}]{pshenay.ivanov.18}%
  \BibitemOpen
  \bibfield  {author} {\bibinfo {author} {\bibfnamefont {D.~A.}\ \bibnamefont
  {Pshenay-Severin}}, \bibinfo {author} {\bibfnamefont {Y.~V.}\ \bibnamefont
  {Ivanov}}, \bibinfo {author} {\bibfnamefont {A.~A.}\ \bibnamefont {Burkov}},\
  and\ \bibinfo {author} {\bibfnamefont {A.~T.}\ \bibnamefont {Burkov}},\
  }\bibfield  {title} {\bibinfo {title} {Band structure and unconventional
  electronic topology of {CoSi}},\ }\href
  {https://doi.org/10.1088/1361-648X/aab0ba} {\bibfield  {journal} {\bibinfo
  {journal} {J. Phys.: Condens. Matter}\ }\textbf {\bibinfo {volume} {30}},\
  \bibinfo {pages} {135501} (\bibinfo {year} {2018})}\BibitemShut {NoStop}%
\bibitem [{\citenamefont {Rao}\ \emph {et~al.}(2019)\citenamefont {Rao},
  \citenamefont {Li}, \citenamefont {Zhang}, \citenamefont {Tian},
  \citenamefont {Li}, \citenamefont {Fu}, \citenamefont {Tang}, \citenamefont
  {Wang}, \citenamefont {Li}, \citenamefont {Fan}, \citenamefont {Li},
  \citenamefont {Huang}, \citenamefont {Liu}, \citenamefont {Long},
  \citenamefont {Fang}, \citenamefont {Weng}, \citenamefont {Shi},
  \citenamefont {Lei}, \citenamefont {Sun}, \citenamefont {Qian},\ and\
  \citenamefont {Ding}}]{rao.li.19}%
  \BibitemOpen
  \bibfield  {author} {\bibinfo {author} {\bibfnamefont {Z.}~\bibnamefont
  {Rao}}, \bibinfo {author} {\bibfnamefont {H.}~\bibnamefont {Li}}, \bibinfo
  {author} {\bibfnamefont {T.}~\bibnamefont {Zhang}}, \bibinfo {author}
  {\bibfnamefont {S.}~\bibnamefont {Tian}}, \bibinfo {author} {\bibfnamefont
  {C.}~\bibnamefont {Li}}, \bibinfo {author} {\bibfnamefont {B.}~\bibnamefont
  {Fu}}, \bibinfo {author} {\bibfnamefont {C.}~\bibnamefont {Tang}}, \bibinfo
  {author} {\bibfnamefont {L.}~\bibnamefont {Wang}}, \bibinfo {author}
  {\bibfnamefont {Z.}~\bibnamefont {Li}}, \bibinfo {author} {\bibfnamefont
  {W.}~\bibnamefont {Fan}}, \bibinfo {author} {\bibfnamefont {J.}~\bibnamefont
  {Li}}, \bibinfo {author} {\bibfnamefont {Y.}~\bibnamefont {Huang}}, \bibinfo
  {author} {\bibfnamefont {Z.}~\bibnamefont {Liu}}, \bibinfo {author}
  {\bibfnamefont {Y.}~\bibnamefont {Long}}, \bibinfo {author} {\bibfnamefont
  {C.}~\bibnamefont {Fang}}, \bibinfo {author} {\bibfnamefont {H.}~\bibnamefont
  {Weng}}, \bibinfo {author} {\bibfnamefont {Y.}~\bibnamefont {Shi}}, \bibinfo
  {author} {\bibfnamefont {H.}~\bibnamefont {Lei}}, \bibinfo {author}
  {\bibfnamefont {Y.}~\bibnamefont {Sun}}, \bibinfo {author} {\bibfnamefont
  {T.}~\bibnamefont {Qian}},\ and\ \bibinfo {author} {\bibfnamefont
  {H.}~\bibnamefont {Ding}},\ }\bibfield  {title} {\bibinfo {title}
  {Observation of unconventional chiral fermions with long fermi arcs in
  {CoSi}},\ }\href {https://doi.org/10.1038/s41586-019-1031-8} {\bibfield
  {journal} {\bibinfo  {journal} {Nature}\ }\textbf {\bibinfo {volume} {567}},\
  \bibinfo {pages} {496} (\bibinfo {year} {2019})}\BibitemShut {NoStop}%
\bibitem [{\citenamefont {Balasubramanian}\ \emph {et~al.}(2020)\citenamefont
  {Balasubramanian}, \citenamefont {Manchanda}, \citenamefont {Pahari},
  \citenamefont {Chen}, \citenamefont {Zhang}, \citenamefont {Valloppilly},
  \citenamefont {Li}, \citenamefont {Sarella}, \citenamefont {Yue},
  \citenamefont {Ullah}, \citenamefont {Dev}, \citenamefont {Muller},
  \citenamefont {Skomski}, \citenamefont {Hadjipanayis},\ and\ \citenamefont
  {Sellmyer}}]{balasubramanian.manchanda.20}%
  \BibitemOpen
  \bibfield  {author} {\bibinfo {author} {\bibfnamefont {B.}~\bibnamefont
  {Balasubramanian}}, \bibinfo {author} {\bibfnamefont {P.}~\bibnamefont
  {Manchanda}}, \bibinfo {author} {\bibfnamefont {R.}~\bibnamefont {Pahari}},
  \bibinfo {author} {\bibfnamefont {Z.}~\bibnamefont {Chen}}, \bibinfo {author}
  {\bibfnamefont {W.}~\bibnamefont {Zhang}}, \bibinfo {author} {\bibfnamefont
  {S.~R.}\ \bibnamefont {Valloppilly}}, \bibinfo {author} {\bibfnamefont
  {X.}~\bibnamefont {Li}}, \bibinfo {author} {\bibfnamefont {A.}~\bibnamefont
  {Sarella}}, \bibinfo {author} {\bibfnamefont {L.}~\bibnamefont {Yue}},
  \bibinfo {author} {\bibfnamefont {A.}~\bibnamefont {Ullah}}, \bibinfo
  {author} {\bibfnamefont {P.}~\bibnamefont {Dev}}, \bibinfo {author}
  {\bibfnamefont {D.~A.}\ \bibnamefont {Muller}}, \bibinfo {author}
  {\bibfnamefont {R.}~\bibnamefont {Skomski}}, \bibinfo {author} {\bibfnamefont
  {G.~C.}\ \bibnamefont {Hadjipanayis}},\ and\ \bibinfo {author} {\bibfnamefont
  {D.~J.}\ \bibnamefont {Sellmyer}},\ }\bibfield  {title} {\bibinfo {title}
  {Chiral magnetism and high-temperature skyrmions in {B20}-ordered {Co-Si}},\
  }\href {https://doi.org/10.1103/PhysRevLett.124.057201} {\bibfield  {journal}
  {\bibinfo  {journal} {Phys. Rev. Lett.}\ }\textbf {\bibinfo {volume} {124}},\
  \bibinfo {pages} {057201} (\bibinfo {year} {2020})}\BibitemShut {NoStop}%
\bibitem [{\citenamefont {Makarova}\ \emph {et~al.}(2012)\citenamefont
  {Makarova}, \citenamefont {Tsvyashchenko}, \citenamefont {Andre},
  \citenamefont {Porcher}, \citenamefont {Fomicheva}, \citenamefont {Rey},\
  and\ \citenamefont {Mirebeau}}]{makarova.tsvyashchenko.12}%
  \BibitemOpen
  \bibfield  {author} {\bibinfo {author} {\bibfnamefont {O.~L.}\ \bibnamefont
  {Makarova}}, \bibinfo {author} {\bibfnamefont {A.~V.}\ \bibnamefont
  {Tsvyashchenko}}, \bibinfo {author} {\bibfnamefont {G.}~\bibnamefont
  {Andre}}, \bibinfo {author} {\bibfnamefont {F.}~\bibnamefont {Porcher}},
  \bibinfo {author} {\bibfnamefont {L.~N.}\ \bibnamefont {Fomicheva}}, \bibinfo
  {author} {\bibfnamefont {N.}~\bibnamefont {Rey}},\ and\ \bibinfo {author}
  {\bibfnamefont {I.}~\bibnamefont {Mirebeau}},\ }\bibfield  {title} {\bibinfo
  {title} {Neutron diffraction study of the chiral magnet {MnGe}},\ }\href
  {https://doi.org/10.1103/PhysRevB.85.205205} {\bibfield  {journal} {\bibinfo
  {journal} {Phys. Rev. B}\ }\textbf {\bibinfo {volume} {85}},\ \bibinfo
  {pages} {205205} (\bibinfo {year} {2012})}\BibitemShut {NoStop}%
\bibitem [{\citenamefont {Kanazawa}\ \emph {et~al.}(2012)\citenamefont
  {Kanazawa}, \citenamefont {Kim}, \citenamefont {Inosov}, \citenamefont
  {White}, \citenamefont {Egetenmeyer}, \citenamefont {Gavilano}, \citenamefont
  {Ishiwata}, \citenamefont {Onose}, \citenamefont {Arima}, \citenamefont
  {Keimer},\ and\ \citenamefont {Tokura}}]{kanazawa.kim.12}%
  \BibitemOpen
  \bibfield  {author} {\bibinfo {author} {\bibfnamefont {N.}~\bibnamefont
  {Kanazawa}}, \bibinfo {author} {\bibfnamefont {J.-H.}\ \bibnamefont {Kim}},
  \bibinfo {author} {\bibfnamefont {D.~S.}\ \bibnamefont {Inosov}}, \bibinfo
  {author} {\bibfnamefont {J.~S.}\ \bibnamefont {White}}, \bibinfo {author}
  {\bibfnamefont {N.}~\bibnamefont {Egetenmeyer}}, \bibinfo {author}
  {\bibfnamefont {J.~L.}\ \bibnamefont {Gavilano}}, \bibinfo {author}
  {\bibfnamefont {S.}~\bibnamefont {Ishiwata}}, \bibinfo {author}
  {\bibfnamefont {Y.}~\bibnamefont {Onose}}, \bibinfo {author} {\bibfnamefont
  {T.}~\bibnamefont {Arima}}, \bibinfo {author} {\bibfnamefont
  {B.}~\bibnamefont {Keimer}},\ and\ \bibinfo {author} {\bibfnamefont
  {Y.}~\bibnamefont {Tokura}},\ }\bibfield  {title} {\bibinfo {title} {Possible
  skyrmion-lattice ground state in the {B20} chiral-lattice magnet {MnGe} as
  seen via small-angle neutron scattering},\ }\href
  {https://doi.org/10.1103/PhysRevB.86.134425} {\bibfield  {journal} {\bibinfo
  {journal} {Phys. Rev. B}\ }\textbf {\bibinfo {volume} {86}},\ \bibinfo
  {pages} {134425} (\bibinfo {year} {2012})}\BibitemShut {NoStop}%
\bibitem [{\citenamefont {Hsieh}\ \emph {et~al.}(2022)\citenamefont {Hsieh},
  \citenamefont {Prasad},\ and\ \citenamefont {Guo}}]{hsieh.prasad.22}%
  \BibitemOpen
  \bibfield  {author} {\bibinfo {author} {\bibfnamefont {T.-Y.}\ \bibnamefont
  {Hsieh}}, \bibinfo {author} {\bibfnamefont {B.~B.}\ \bibnamefont {Prasad}},\
  and\ \bibinfo {author} {\bibfnamefont {G.-Y.}\ \bibnamefont {Guo}},\
  }\bibfield  {title} {\bibinfo {title} {Helicity-tunable spin hall and spin
  {Nernst} effects in unconventional chiral fermion semimetals {$XY$} ({$X=$
  Co, Rh}; {$Y=$ Si, Ge})},\ }\href
  {https://doi.org/10.1103/PhysRevB.106.165102} {\bibfield  {journal} {\bibinfo
   {journal} {Phys. Rev. B}\ }\textbf {\bibinfo {volume} {106}},\ \bibinfo
  {pages} {165102} (\bibinfo {year} {2022})}\BibitemShut {NoStop}%
\bibitem [{\citenamefont {Pfau}\ \emph {et~al.}(2019)\citenamefont {Pfau},
  \citenamefont {{Trey Diulus}}, \citenamefont {He}, \citenamefont
  {Albuquerque}, \citenamefont {Stickle},\ and\ \citenamefont
  {Herman}}]{pfau.trey.19}%
  \BibitemOpen
  \bibfield  {author} {\bibinfo {author} {\bibfnamefont {A.~J.}\ \bibnamefont
  {Pfau}}, \bibinfo {author} {\bibfnamefont {J.}~\bibnamefont {{Trey Diulus}}},
  \bibinfo {author} {\bibfnamefont {S.}~\bibnamefont {He}}, \bibinfo {author}
  {\bibfnamefont {G.~H.}\ \bibnamefont {Albuquerque}}, \bibinfo {author}
  {\bibfnamefont {W.~F.}\ \bibnamefont {Stickle}},\ and\ \bibinfo {author}
  {\bibfnamefont {G.~S.}\ \bibnamefont {Herman}},\ }\bibfield  {title}
  {\bibinfo {title} {{CoGe} surface oxidation studied using {X}-ray
  photoelectron spectroscopy},\ }\href
  {https://doi.org/10.1016/j.apsusc.2018.11.019} {\bibfield  {journal}
  {\bibinfo  {journal} {Appl. Surf. Sci.}\ }\textbf {\bibinfo {volume} {469}},\
  \bibinfo {pages} {298} (\bibinfo {year} {2019})}\BibitemShut {NoStop}%
\bibitem [{\citenamefont {K\'uko\'lov\'a}\ \emph {et~al.}(2021)\citenamefont
  {K\'uko\'lov\'a}, \citenamefont {Dimitrievska}, \citenamefont {Litvinchuk},
  \citenamefont {Ramanandan}, \citenamefont {Tappy}, \citenamefont {Menon},
  \citenamefont {Borg}, \citenamefont {Grundler},\ and\ \citenamefont
  {Fontcuberta~i Morral}}]{kukolova.dimitrievska.21}%
  \BibitemOpen
  \bibfield  {author} {\bibinfo {author} {\bibfnamefont {A.}~\bibnamefont
  {K\'uko\'lov\'a}}, \bibinfo {author} {\bibfnamefont {M.}~\bibnamefont
  {Dimitrievska}}, \bibinfo {author} {\bibfnamefont {A.~P.}\ \bibnamefont
  {Litvinchuk}}, \bibinfo {author} {\bibfnamefont {S.~P.}\ \bibnamefont
  {Ramanandan}}, \bibinfo {author} {\bibfnamefont {N.}~\bibnamefont {Tappy}},
  \bibinfo {author} {\bibfnamefont {H.}~\bibnamefont {Menon}}, \bibinfo
  {author} {\bibfnamefont {M.}~\bibnamefont {Borg}}, \bibinfo {author}
  {\bibfnamefont {D.}~\bibnamefont {Grundler}},\ and\ \bibinfo {author}
  {\bibfnamefont {A.}~\bibnamefont {Fontcuberta~i Morral}},\ }\bibfield
  {title} {\bibinfo {title} {Cubic, hexagonal and tetragonal {FeGe$_{x}$}
  phases ($x = 1, 1.5, 2$): {Raman} spectroscopy and magnetic properties},\
  }\href {https://doi.org/10.1039/D1CE00970B} {\bibfield  {journal} {\bibinfo
  {journal} {CrystEngComm}\ }\textbf {\bibinfo {volume} {23}},\ \bibinfo
  {pages} {6506} (\bibinfo {year} {2021})}\BibitemShut {NoStop}%
\bibitem [{\citenamefont {Ptok}\ \emph {et~al.}(2021)\citenamefont {Ptok},
  \citenamefont {Kobia\l{}ka}, \citenamefont {Sternik}, \citenamefont
  {\L{}a\.{z}ewski}, \citenamefont {Jochym}, \citenamefont {Ole\'{s}},
  \citenamefont {Stankov},\ and\ \citenamefont {Piekarz}}]{ptok.kobialka.21}%
  \BibitemOpen
  \bibfield  {author} {\bibinfo {author} {\bibfnamefont {A.}~\bibnamefont
  {Ptok}}, \bibinfo {author} {\bibfnamefont {A.}~\bibnamefont {Kobia\l{}ka}},
  \bibinfo {author} {\bibfnamefont {M.}~\bibnamefont {Sternik}}, \bibinfo
  {author} {\bibfnamefont {J.}~\bibnamefont {\L{}a\.{z}ewski}}, \bibinfo
  {author} {\bibfnamefont {P.~T.}\ \bibnamefont {Jochym}}, \bibinfo {author}
  {\bibfnamefont {A.~M.}\ \bibnamefont {Ole\'{s}}}, \bibinfo {author}
  {\bibfnamefont {S.}~\bibnamefont {Stankov}},\ and\ \bibinfo {author}
  {\bibfnamefont {P.}~\bibnamefont {Piekarz}},\ }\bibfield  {title} {\bibinfo
  {title} {Chiral phonons in the honeycomb sublattice of layered {CoSn}-like
  compounds},\ }\href {https://doi.org/10.1103/PhysRevB.104.054305} {\bibfield
  {journal} {\bibinfo  {journal} {Phys. Rev. B}\ }\textbf {\bibinfo {volume}
  {104}},\ \bibinfo {pages} {054305} (\bibinfo {year} {2021})}\BibitemShut
  {NoStop}%
\bibitem [{\citenamefont {Saal}\ \emph {et~al.}(2013)\citenamefont {Saal},
  \citenamefont {Kirklin}, \citenamefont {Aykol}, \citenamefont {Meredig},\
  and\ \citenamefont {Wolverton}}]{saal.kirklin.13}%
  \BibitemOpen
  \bibfield  {author} {\bibinfo {author} {\bibfnamefont {J.~E.}\ \bibnamefont
  {Saal}}, \bibinfo {author} {\bibfnamefont {S.}~\bibnamefont {Kirklin}},
  \bibinfo {author} {\bibfnamefont {M.}~\bibnamefont {Aykol}}, \bibinfo
  {author} {\bibfnamefont {B.}~\bibnamefont {Meredig}},\ and\ \bibinfo {author}
  {\bibfnamefont {C.}~\bibnamefont {Wolverton}},\ }\bibfield  {title} {\bibinfo
  {title} {Materials design and discovery with high-throughput density
  functional theory: The open quantum materials database ({OQMD})},\ }\href
  {https://doi.org/10.1007/s11837-013-0755-4} {\bibfield  {journal} {\bibinfo
  {journal} {JOM}\ }\textbf {\bibinfo {volume} {65}},\ \bibinfo {pages} {1501}
  (\bibinfo {year} {2013})}\BibitemShut {NoStop}%
\bibitem [{\citenamefont {Kirklin}\ \emph {et~al.}(2015)\citenamefont
  {Kirklin}, \citenamefont {Saal}, \citenamefont {Meredig}, \citenamefont
  {Thompson}, \citenamefont {Doak}, \citenamefont {Aykol}, \citenamefont
  {R{\"u}hl},\ and\ \citenamefont {Wolverton}}]{kirklin.saal.15}%
  \BibitemOpen
  \bibfield  {author} {\bibinfo {author} {\bibfnamefont {S.}~\bibnamefont
  {Kirklin}}, \bibinfo {author} {\bibfnamefont {J.~E.}\ \bibnamefont {Saal}},
  \bibinfo {author} {\bibfnamefont {B.}~\bibnamefont {Meredig}}, \bibinfo
  {author} {\bibfnamefont {A.}~\bibnamefont {Thompson}}, \bibinfo {author}
  {\bibfnamefont {J.~W.}\ \bibnamefont {Doak}}, \bibinfo {author}
  {\bibfnamefont {M.}~\bibnamefont {Aykol}}, \bibinfo {author} {\bibfnamefont
  {S.}~\bibnamefont {R{\"u}hl}},\ and\ \bibinfo {author} {\bibfnamefont
  {C.}~\bibnamefont {Wolverton}},\ }\bibfield  {title} {\bibinfo {title} {The
  open quantum materials database ({OQMD}): assessing the accuracy of {DFT}
  formation energies},\ }\href {https://doi.org/10.1038/npjcompumats.2015.10}
  {\bibfield  {journal} {\bibinfo  {journal} {npj Comput. Mater.}\ }\textbf
  {\bibinfo {volume} {1}},\ \bibinfo {pages} {15010} (\bibinfo {year}
  {2015})}\BibitemShut {NoStop}%
\bibitem [{Note1()}]{Note1}%
  \BibitemOpen
  \bibinfo {note} {\protect \url
  {https://oqmd.org/materials/composition/CoGe}}\BibitemShut {NoStop}%
\bibitem [{\citenamefont {Richardson}(1967{\natexlab{b}})}]{richardson.67b}%
  \BibitemOpen
  \bibfield  {author} {\bibinfo {author} {\bibfnamefont {M.~W.}\ \bibnamefont
  {Richardson}},\ }\bibfield  {title} {\bibinfo {title} {The partial
  equilibrium diagram of the {Fe-Ge} system in the range 40-72 at. \% {Ge}, and
  the crystallisation of some iron germanides by chemical transport
  reactions},\ }\href {https://doi.org/10.3891/acta.chem.scand.21-2305}
  {\bibfield  {journal} {\bibinfo  {journal} {Acta Chemica Scandinavica}\
  }\textbf {\bibinfo {volume} {21}},\ \bibinfo {pages} {2305} (\bibinfo {year}
  {1967}{\natexlab{b}})}\BibitemShut {NoStop}%
\bibitem [{\citenamefont {Larsson}\ \emph {et~al.}(1996)\citenamefont
  {Larsson}, \citenamefont {Haeberlein}, \citenamefont {Lidin},\ and\
  \citenamefont {Schwarz}}]{larsson.haeberlein.96}%
  \BibitemOpen
  \bibfield  {author} {\bibinfo {author} {\bibfnamefont {A.}~\bibnamefont
  {Larsson}}, \bibinfo {author} {\bibfnamefont {M.}~\bibnamefont {Haeberlein}},
  \bibinfo {author} {\bibfnamefont {S.}~\bibnamefont {Lidin}},\ and\ \bibinfo
  {author} {\bibfnamefont {U.}~\bibnamefont {Schwarz}},\ }\bibfield  {title}
  {\bibinfo {title} {Single crystal structure refinement and high-pressure
  properties of {CoSn}},\ }\href {https://doi.org/10.1016/0925-8388(95)02189-2}
  {\bibfield  {journal} {\bibinfo  {journal} {J. Alloys Compd.}\ }\textbf
  {\bibinfo {volume} {240}},\ \bibinfo {pages} {79} (\bibinfo {year}
  {1996})}\BibitemShut {NoStop}%
\bibitem [{\citenamefont {Waerenborgh}\ \emph {et~al.}(2005)\citenamefont
  {Waerenborgh}, \citenamefont {Pereira}, \citenamefont {Gonçalves},\ and\
  \citenamefont {No\"{e}l}}]{waerenborgh.pereira.05}%
  \BibitemOpen
  \bibfield  {author} {\bibinfo {author} {\bibfnamefont {J.}~\bibnamefont
  {Waerenborgh}}, \bibinfo {author} {\bibfnamefont {L.}~\bibnamefont
  {Pereira}}, \bibinfo {author} {\bibfnamefont {A.}~\bibnamefont
  {Gonçalves}},\ and\ \bibinfo {author} {\bibfnamefont {H.}~\bibnamefont
  {No\"{e}l}},\ }\bibfield  {title} {\bibinfo {title} {Crystal structure,
  {$^{57}$Fe} {M\"{o}ssbauer} spectroscopy and magnetization of
  {U$_{x}$Fe$_{6}$Sn$_{6}$} ($0 \leq x \leq 0.6$)},\ }\href
  {https://doi.org/https://doi.org/10.1016/j.intermet.2004.08.009} {\bibfield
  {journal} {\bibinfo  {journal} {Intermetallics}\ }\textbf {\bibinfo {volume}
  {13}},\ \bibinfo {pages} {490} (\bibinfo {year} {2005})}\BibitemShut
  {NoStop}%
\bibitem [{\citenamefont {Pfisterer}\ and\ \citenamefont
  {Schubert}(1950)}]{pfisterer.schubert.50}%
  \BibitemOpen
  \bibfield  {author} {\bibinfo {author} {\bibfnamefont {H.}~\bibnamefont
  {Pfisterer}}\ and\ \bibinfo {author} {\bibfnamefont {K.}~\bibnamefont
  {Schubert}},\ }\bibfield  {title} {\bibinfo {title} {Neue phasen vom {MnP}
  ({B31})-typ},\ }\href {https://doi.org/10.1007/BF00623719} {\bibfield
  {journal} {\bibinfo  {journal} {Naturwissenschaften}\ }\textbf {\bibinfo
  {volume} {37}},\ \bibinfo {pages} {112} (\bibinfo {year} {1950})}\BibitemShut
  {NoStop}%
\bibitem [{\citenamefont {Geller}(1955)}]{geller.55}%
  \BibitemOpen
  \bibfield  {author} {\bibinfo {author} {\bibfnamefont {S.}~\bibnamefont
  {Geller}},\ }\bibfield  {title} {\bibinfo {title} {The rhodium--germanium
  system. {I}. the crystal structures of {Rh$_{2}$Ge}, {Rh$_{5}$Ge$_{3}$} and
  {RhGe}},\ }\href {https://doi.org/10.1107/S0365110X55000030} {\bibfield
  {journal} {\bibinfo  {journal} {Acta Crystallographica}\ }\textbf {\bibinfo
  {volume} {8}},\ \bibinfo {pages} {15} (\bibinfo {year} {1955})}\BibitemShut
  {NoStop}%
\bibitem [{\citenamefont {Wang}\ \emph {et~al.}(2021)\citenamefont {Wang},
  \citenamefont {Botti},\ and\ \citenamefont {Marques}}]{wang.botti.21}%
  \BibitemOpen
  \bibfield  {author} {\bibinfo {author} {\bibfnamefont {H.-C.}\ \bibnamefont
  {Wang}}, \bibinfo {author} {\bibfnamefont {S.}~\bibnamefont {Botti}},\ and\
  \bibinfo {author} {\bibfnamefont {M.~A.~L.}\ \bibnamefont {Marques}},\
  }\bibfield  {title} {\bibinfo {title} {Predicting stable crystalline
  compounds using chemical similarity},\ }\href
  {https://doi.org/10.1038/s41524-020-00481-6} {\bibfield  {journal} {\bibinfo
  {journal} {npj Comput. Mater.}\ }\textbf {\bibinfo {volume} {7}},\ \bibinfo
  {pages} {12} (\bibinfo {year} {2021})}\BibitemShut {NoStop}%
\bibitem [{\citenamefont {Sales}\ \emph {et~al.}(2019)\citenamefont {Sales},
  \citenamefont {Yan}, \citenamefont {Meier}, \citenamefont {Christianson},
  \citenamefont {Okamoto},\ and\ \citenamefont {McGuire}}]{sales.yan.19}%
  \BibitemOpen
  \bibfield  {author} {\bibinfo {author} {\bibfnamefont {B.~C.}\ \bibnamefont
  {Sales}}, \bibinfo {author} {\bibfnamefont {J.}~\bibnamefont {Yan}}, \bibinfo
  {author} {\bibfnamefont {W.~R.}\ \bibnamefont {Meier}}, \bibinfo {author}
  {\bibfnamefont {A.~D.}\ \bibnamefont {Christianson}}, \bibinfo {author}
  {\bibfnamefont {S.}~\bibnamefont {Okamoto}},\ and\ \bibinfo {author}
  {\bibfnamefont {M.~A.}\ \bibnamefont {McGuire}},\ }\bibfield  {title}
  {\bibinfo {title} {Electronic, magnetic, and thermodynamic properties of the
  kagome layer compound {FeSn}},\ }\href
  {https://doi.org/10.1103/PhysRevMaterials.3.114203} {\bibfield  {journal}
  {\bibinfo  {journal} {Phys. Rev. Materials}\ }\textbf {\bibinfo {volume}
  {3}},\ \bibinfo {pages} {114203} (\bibinfo {year} {2019})}\BibitemShut
  {NoStop}%
\bibitem [{\citenamefont {Meier}\ \emph {et~al.}(2020)\citenamefont {Meier},
  \citenamefont {Du}, \citenamefont {Okamoto}, \citenamefont {Mohanta},
  \citenamefont {May}, \citenamefont {McGuire}, \citenamefont {Bridges},
  \citenamefont {Samolyuk},\ and\ \citenamefont {Sales}}]{meier.du.20}%
  \BibitemOpen
  \bibfield  {author} {\bibinfo {author} {\bibfnamefont {W.~R.}\ \bibnamefont
  {Meier}}, \bibinfo {author} {\bibfnamefont {M.-H.}\ \bibnamefont {Du}},
  \bibinfo {author} {\bibfnamefont {S.}~\bibnamefont {Okamoto}}, \bibinfo
  {author} {\bibfnamefont {N.}~\bibnamefont {Mohanta}}, \bibinfo {author}
  {\bibfnamefont {A.~F.}\ \bibnamefont {May}}, \bibinfo {author} {\bibfnamefont
  {M.~A.}\ \bibnamefont {McGuire}}, \bibinfo {author} {\bibfnamefont {C.~A.}\
  \bibnamefont {Bridges}}, \bibinfo {author} {\bibfnamefont {G.~D.}\
  \bibnamefont {Samolyuk}},\ and\ \bibinfo {author} {\bibfnamefont {B.~C.}\
  \bibnamefont {Sales}},\ }\bibfield  {title} {\bibinfo {title} {Flat bands in
  the {CoSn}-type compounds},\ }\href
  {https://doi.org/10.1103/PhysRevB.102.075148} {\bibfield  {journal} {\bibinfo
   {journal} {Phys. Rev. B}\ }\textbf {\bibinfo {volume} {102}},\ \bibinfo
  {pages} {075148} (\bibinfo {year} {2020})}\BibitemShut {NoStop}%
\bibitem [{\citenamefont {Kang}\ \emph
  {et~al.}(2020{\natexlab{a}})\citenamefont {Kang}, \citenamefont {Fang},
  \citenamefont {Ye}, \citenamefont {Po}, \citenamefont {Denlinger},
  \citenamefont {Jozwiak}, \citenamefont {Bostwick}, \citenamefont {Rotenberg},
  \citenamefont {Kaxiras}, \citenamefont {Checkelsky},\ and\ \citenamefont
  {Comin}}]{kang.fang.20}%
  \BibitemOpen
  \bibfield  {author} {\bibinfo {author} {\bibfnamefont {M.}~\bibnamefont
  {Kang}}, \bibinfo {author} {\bibfnamefont {S.}~\bibnamefont {Fang}}, \bibinfo
  {author} {\bibfnamefont {L.}~\bibnamefont {Ye}}, \bibinfo {author}
  {\bibfnamefont {H.~C.}\ \bibnamefont {Po}}, \bibinfo {author} {\bibfnamefont
  {J.}~\bibnamefont {Denlinger}}, \bibinfo {author} {\bibfnamefont
  {C.}~\bibnamefont {Jozwiak}}, \bibinfo {author} {\bibfnamefont
  {A.}~\bibnamefont {Bostwick}}, \bibinfo {author} {\bibfnamefont
  {E.}~\bibnamefont {Rotenberg}}, \bibinfo {author} {\bibfnamefont
  {E.}~\bibnamefont {Kaxiras}}, \bibinfo {author} {\bibfnamefont {J.~G.}\
  \bibnamefont {Checkelsky}},\ and\ \bibinfo {author} {\bibfnamefont
  {R.}~\bibnamefont {Comin}},\ }\bibfield  {title} {\bibinfo {title}
  {Topological flat bands in frustrated kagome lattice {CoSn}},\ }\href
  {https://doi.org/10.1038/s41467-020-17465-1} {\bibfield  {journal} {\bibinfo
  {journal} {Nat. Commun.}\ }\textbf {\bibinfo {volume} {11}},\ \bibinfo
  {pages} {4004} (\bibinfo {year} {2020}{\natexlab{a}})}\BibitemShut {NoStop}%
\bibitem [{\citenamefont {Kang}\ \emph
  {et~al.}(2020{\natexlab{b}})\citenamefont {Kang}, \citenamefont {Ye},
  \citenamefont {Fang}, \citenamefont {You}, \citenamefont {Levitan},
  \citenamefont {Han}, \citenamefont {Facio}, \citenamefont {Jozwiak},
  \citenamefont {Bostwick}, \citenamefont {Rotenberg}, \citenamefont {Chan},
  \citenamefont {McDonald}, \citenamefont {Graf}, \citenamefont {Kaznatcheev},
  \citenamefont {Vescovo}, \citenamefont {Bell}, \citenamefont {Kaxiras},
  \citenamefont {van~den Brink}, \citenamefont {Richter}, \citenamefont
  {Prasad~Ghimire}, \citenamefont {Checkelsky},\ and\ \citenamefont
  {Comin}}]{kang.ye.20}%
  \BibitemOpen
  \bibfield  {author} {\bibinfo {author} {\bibfnamefont {M.}~\bibnamefont
  {Kang}}, \bibinfo {author} {\bibfnamefont {L.}~\bibnamefont {Ye}}, \bibinfo
  {author} {\bibfnamefont {S.}~\bibnamefont {Fang}}, \bibinfo {author}
  {\bibfnamefont {J.-S.}\ \bibnamefont {You}}, \bibinfo {author} {\bibfnamefont
  {A.}~\bibnamefont {Levitan}}, \bibinfo {author} {\bibfnamefont
  {M.}~\bibnamefont {Han}}, \bibinfo {author} {\bibfnamefont {J.~I.}\
  \bibnamefont {Facio}}, \bibinfo {author} {\bibfnamefont {C.}~\bibnamefont
  {Jozwiak}}, \bibinfo {author} {\bibfnamefont {A.}~\bibnamefont {Bostwick}},
  \bibinfo {author} {\bibfnamefont {E.}~\bibnamefont {Rotenberg}}, \bibinfo
  {author} {\bibfnamefont {M.~K.}\ \bibnamefont {Chan}}, \bibinfo {author}
  {\bibfnamefont {R.~D.}\ \bibnamefont {McDonald}}, \bibinfo {author}
  {\bibfnamefont {D.}~\bibnamefont {Graf}}, \bibinfo {author} {\bibfnamefont
  {K.}~\bibnamefont {Kaznatcheev}}, \bibinfo {author} {\bibfnamefont
  {E.}~\bibnamefont {Vescovo}}, \bibinfo {author} {\bibfnamefont {D.~C.}\
  \bibnamefont {Bell}}, \bibinfo {author} {\bibfnamefont {E.}~\bibnamefont
  {Kaxiras}}, \bibinfo {author} {\bibfnamefont {J.}~\bibnamefont {van~den
  Brink}}, \bibinfo {author} {\bibfnamefont {M.}~\bibnamefont {Richter}},
  \bibinfo {author} {\bibfnamefont {M.}~\bibnamefont {Prasad~Ghimire}},
  \bibinfo {author} {\bibfnamefont {J.~G.}\ \bibnamefont {Checkelsky}},\ and\
  \bibinfo {author} {\bibfnamefont {R.}~\bibnamefont {Comin}},\ }\bibfield
  {title} {\bibinfo {title} {Dirac fermions and flat bands in the ideal kagome
  metal {FeSn}},\ }\href {https://doi.org/10.1038/s41563-019-0531-0} {\bibfield
   {journal} {\bibinfo  {journal} {Nat. Mater.}\ }\textbf {\bibinfo {volume}
  {19}},\ \bibinfo {pages} {163} (\bibinfo {year}
  {2020}{\natexlab{b}})}\BibitemShut {NoStop}%
\bibitem [{\citenamefont {Lin}\ \emph {et~al.}(2020)\citenamefont {Lin},
  \citenamefont {Wang}, \citenamefont {Wang}, \citenamefont {Yi}, \citenamefont
  {Li}, \citenamefont {Zhang}, \citenamefont {Wang}, \citenamefont {Wang},
  \citenamefont {Huang}, \citenamefont {Sun}, \citenamefont {Huang},
  \citenamefont {Shen}, \citenamefont {Feng}, \citenamefont {Sun},
  \citenamefont {Cho}, \citenamefont {Zeng},\ and\ \citenamefont
  {Zhang}}]{lin.wang.20}%
  \BibitemOpen
  \bibfield  {author} {\bibinfo {author} {\bibfnamefont {Z.}~\bibnamefont
  {Lin}}, \bibinfo {author} {\bibfnamefont {C.}~\bibnamefont {Wang}}, \bibinfo
  {author} {\bibfnamefont {P.}~\bibnamefont {Wang}}, \bibinfo {author}
  {\bibfnamefont {S.}~\bibnamefont {Yi}}, \bibinfo {author} {\bibfnamefont
  {L.}~\bibnamefont {Li}}, \bibinfo {author} {\bibfnamefont {Q.}~\bibnamefont
  {Zhang}}, \bibinfo {author} {\bibfnamefont {Y.}~\bibnamefont {Wang}},
  \bibinfo {author} {\bibfnamefont {Z.}~\bibnamefont {Wang}}, \bibinfo {author}
  {\bibfnamefont {H.}~\bibnamefont {Huang}}, \bibinfo {author} {\bibfnamefont
  {Y.}~\bibnamefont {Sun}}, \bibinfo {author} {\bibfnamefont {Y.}~\bibnamefont
  {Huang}}, \bibinfo {author} {\bibfnamefont {D.}~\bibnamefont {Shen}},
  \bibinfo {author} {\bibfnamefont {D.}~\bibnamefont {Feng}}, \bibinfo {author}
  {\bibfnamefont {Z.}~\bibnamefont {Sun}}, \bibinfo {author} {\bibfnamefont
  {J.-H.}\ \bibnamefont {Cho}}, \bibinfo {author} {\bibfnamefont
  {C.}~\bibnamefont {Zeng}},\ and\ \bibinfo {author} {\bibfnamefont
  {Z.}~\bibnamefont {Zhang}},\ }\bibfield  {title} {\bibinfo {title} {Dirac
  fermions in antiferromagnetic {FeSn} kagome lattices with combined space
  inversion and time-reversal symmetry},\ }\href
  {https://doi.org/10.1103/PhysRevB.102.155103} {\bibfield  {journal} {\bibinfo
   {journal} {Phys. Rev. B}\ }\textbf {\bibinfo {volume} {102}},\ \bibinfo
  {pages} {155103} (\bibinfo {year} {2020})}\BibitemShut {NoStop}%
\bibitem [{\citenamefont {Huang}\ \emph {et~al.}(2022)\citenamefont {Huang},
  \citenamefont {Zheng}, \citenamefont {Lin}, \citenamefont {Guo},
  \citenamefont {Wang}, \citenamefont {Zhang}, \citenamefont {Zhang},
  \citenamefont {Sun}, \citenamefont {Wang}, \citenamefont {Weng},
  \citenamefont {Li}, \citenamefont {Wu}, \citenamefont {Chen},\ and\
  \citenamefont {Zeng}}]{huang.zheng.22}%
  \BibitemOpen
  \bibfield  {author} {\bibinfo {author} {\bibfnamefont {H.}~\bibnamefont
  {Huang}}, \bibinfo {author} {\bibfnamefont {L.}~\bibnamefont {Zheng}},
  \bibinfo {author} {\bibfnamefont {Z.}~\bibnamefont {Lin}}, \bibinfo {author}
  {\bibfnamefont {X.}~\bibnamefont {Guo}}, \bibinfo {author} {\bibfnamefont
  {S.}~\bibnamefont {Wang}}, \bibinfo {author} {\bibfnamefont {S.}~\bibnamefont
  {Zhang}}, \bibinfo {author} {\bibfnamefont {C.}~\bibnamefont {Zhang}},
  \bibinfo {author} {\bibfnamefont {Z.}~\bibnamefont {Sun}}, \bibinfo {author}
  {\bibfnamefont {Z.}~\bibnamefont {Wang}}, \bibinfo {author} {\bibfnamefont
  {H.}~\bibnamefont {Weng}}, \bibinfo {author} {\bibfnamefont {L.}~\bibnamefont
  {Li}}, \bibinfo {author} {\bibfnamefont {T.}~\bibnamefont {Wu}}, \bibinfo
  {author} {\bibfnamefont {X.}~\bibnamefont {Chen}},\ and\ \bibinfo {author}
  {\bibfnamefont {C.}~\bibnamefont {Zeng}},\ }\bibfield  {title} {\bibinfo
  {title} {Flat-band-induced anomalous anisotropic charge transport and orbital
  magnetism in kagome metal {CoSn}},\ }\href
  {https://doi.org/10.1103/PhysRevLett.128.096601} {\bibfield  {journal}
  {\bibinfo  {journal} {Phys. Rev. Lett.}\ }\textbf {\bibinfo {volume} {128}},\
  \bibinfo {pages} {096601} (\bibinfo {year} {2022})}\BibitemShut {NoStop}%
\bibitem [{\citenamefont {Liu}\ \emph {et~al.}(2020)\citenamefont {Liu},
  \citenamefont {Li}, \citenamefont {Wang}, \citenamefont {Wang}, \citenamefont
  {Wen}, \citenamefont {Jiang}, \citenamefont {Lu}, \citenamefont {Yan},
  \citenamefont {Huang}, \citenamefont {Shen}, \citenamefont {Yin},
  \citenamefont {Wang}, \citenamefont {Yin}, \citenamefont {Lei},\ and\
  \citenamefont {Wang}}]{liu.li.20}%
  \BibitemOpen
  \bibfield  {author} {\bibinfo {author} {\bibfnamefont {Z.}~\bibnamefont
  {Liu}}, \bibinfo {author} {\bibfnamefont {M.}~\bibnamefont {Li}}, \bibinfo
  {author} {\bibfnamefont {Q.}~\bibnamefont {Wang}}, \bibinfo {author}
  {\bibfnamefont {G.}~\bibnamefont {Wang}}, \bibinfo {author} {\bibfnamefont
  {C.}~\bibnamefont {Wen}}, \bibinfo {author} {\bibfnamefont {K.}~\bibnamefont
  {Jiang}}, \bibinfo {author} {\bibfnamefont {X.}~\bibnamefont {Lu}}, \bibinfo
  {author} {\bibfnamefont {S.}~\bibnamefont {Yan}}, \bibinfo {author}
  {\bibfnamefont {Y.}~\bibnamefont {Huang}}, \bibinfo {author} {\bibfnamefont
  {D.}~\bibnamefont {Shen}}, \bibinfo {author} {\bibfnamefont {J.-X.}\
  \bibnamefont {Yin}}, \bibinfo {author} {\bibfnamefont {Z.}~\bibnamefont
  {Wang}}, \bibinfo {author} {\bibfnamefont {Z.}~\bibnamefont {Yin}}, \bibinfo
  {author} {\bibfnamefont {H.}~\bibnamefont {Lei}},\ and\ \bibinfo {author}
  {\bibfnamefont {S.}~\bibnamefont {Wang}},\ }\bibfield  {title} {\bibinfo
  {title} {Orbital-selective {Dirac} fermions and extremely flat bands in
  frustrated kagome-lattice metal {CoSn}},\ }\href
  {https://doi.org/10.1038/s41467-020-17462-4} {\bibfield  {journal} {\bibinfo
  {journal} {Nat. Commun.}\ }\textbf {\bibinfo {volume} {11}},\ \bibinfo
  {pages} {4002} (\bibinfo {year} {2020})}\BibitemShut {NoStop}%
\bibitem [{\citenamefont {Han}\ \emph {et~al.}(2021)\citenamefont {Han},
  \citenamefont {Inoue}, \citenamefont {Fang}, \citenamefont {John},
  \citenamefont {Ye}, \citenamefont {Chan}, \citenamefont {Graf}, \citenamefont
  {Suzuki}, \citenamefont {Ghimire}, \citenamefont {Cho}, \citenamefont
  {Kaxiras},\ and\ \citenamefont {Checkelsky}}]{han.inoue.21}%
  \BibitemOpen
  \bibfield  {author} {\bibinfo {author} {\bibfnamefont {M.}~\bibnamefont
  {Han}}, \bibinfo {author} {\bibfnamefont {H.}~\bibnamefont {Inoue}}, \bibinfo
  {author} {\bibfnamefont {S.}~\bibnamefont {Fang}}, \bibinfo {author}
  {\bibfnamefont {C.}~\bibnamefont {John}}, \bibinfo {author} {\bibfnamefont
  {L.}~\bibnamefont {Ye}}, \bibinfo {author} {\bibfnamefont {M.~K.}\
  \bibnamefont {Chan}}, \bibinfo {author} {\bibfnamefont {D.}~\bibnamefont
  {Graf}}, \bibinfo {author} {\bibfnamefont {T.}~\bibnamefont {Suzuki}},
  \bibinfo {author} {\bibfnamefont {M.~P.}\ \bibnamefont {Ghimire}}, \bibinfo
  {author} {\bibfnamefont {W.~J.}\ \bibnamefont {Cho}}, \bibinfo {author}
  {\bibfnamefont {E.}~\bibnamefont {Kaxiras}},\ and\ \bibinfo {author}
  {\bibfnamefont {J.~G.}\ \bibnamefont {Checkelsky}},\ }\bibfield  {title}
  {\bibinfo {title} {Evidence of two-dimensional flat band at the surface of
  antiferromagnetic kagome metal {FeSn}},\ }\href
  {https://doi.org/10.1038/s41467-021-25705-1} {\bibfield  {journal} {\bibinfo
  {journal} {Nat. Commun.}\ }\textbf {\bibinfo {volume} {12}},\ \bibinfo
  {pages} {5345} (\bibinfo {year} {2021})}\BibitemShut {NoStop}%
\bibitem [{\citenamefont {Sales}\ \emph {et~al.}(2021)\citenamefont {Sales},
  \citenamefont {Meier}, \citenamefont {May}, \citenamefont {Xing},
  \citenamefont {Yan}, \citenamefont {Gao}, \citenamefont {Liu}, \citenamefont
  {Stone}, \citenamefont {Christianson}, \citenamefont {Zhang},\ and\
  \citenamefont {McGuire}}]{sales.meier.21}%
  \BibitemOpen
  \bibfield  {author} {\bibinfo {author} {\bibfnamefont {B.~C.}\ \bibnamefont
  {Sales}}, \bibinfo {author} {\bibfnamefont {W.~R.}\ \bibnamefont {Meier}},
  \bibinfo {author} {\bibfnamefont {A.~F.}\ \bibnamefont {May}}, \bibinfo
  {author} {\bibfnamefont {J.}~\bibnamefont {Xing}}, \bibinfo {author}
  {\bibfnamefont {J.-Q.}\ \bibnamefont {Yan}}, \bibinfo {author} {\bibfnamefont
  {S.}~\bibnamefont {Gao}}, \bibinfo {author} {\bibfnamefont {Y.~H.}\
  \bibnamefont {Liu}}, \bibinfo {author} {\bibfnamefont {M.~B.}\ \bibnamefont
  {Stone}}, \bibinfo {author} {\bibfnamefont {A.~D.}\ \bibnamefont
  {Christianson}}, \bibinfo {author} {\bibfnamefont {Q.}~\bibnamefont
  {Zhang}},\ and\ \bibinfo {author} {\bibfnamefont {M.~A.}\ \bibnamefont
  {McGuire}},\ }\bibfield  {title} {\bibinfo {title} {Tuning the flat bands of
  the kagome metal {CoSn} with {Fe}, {In}, or {Ni} doping},\ }\href
  {https://doi.org/10.1103/PhysRevMaterials.5.044202} {\bibfield  {journal}
  {\bibinfo  {journal} {Phys. Rev. Materials}\ }\textbf {\bibinfo {volume}
  {5}},\ \bibinfo {pages} {044202} (\bibinfo {year} {2021})}\BibitemShut
  {NoStop}%
\bibitem [{\citenamefont {Wu}\ \emph {et~al.}(2015{\natexlab{a}})\citenamefont
  {Wu}, \citenamefont {Yang}, \citenamefont {Le}, \citenamefont {Fan},\ and\
  \citenamefont {Hu}}]{wu.yang.15}%
  \BibitemOpen
  \bibfield  {author} {\bibinfo {author} {\bibfnamefont {X.}~\bibnamefont
  {Wu}}, \bibinfo {author} {\bibfnamefont {F.}~\bibnamefont {Yang}}, \bibinfo
  {author} {\bibfnamefont {C.}~\bibnamefont {Le}}, \bibinfo {author}
  {\bibfnamefont {H.}~\bibnamefont {Fan}},\ and\ \bibinfo {author}
  {\bibfnamefont {J.}~\bibnamefont {Hu}},\ }\bibfield  {title} {\bibinfo
  {title} {Triplet ${p}_{z}$-wave pairing in quasi-one-dimensional
  {${A}_{2}$Cr$_{3}$As$_{3}$} superconductors ({$A=$K, Rb, Cs})},\ }\href
  {https://doi.org/10.1103/PhysRevB.92.104511} {\bibfield  {journal} {\bibinfo
  {journal} {Phys. Rev. B}\ }\textbf {\bibinfo {volume} {92}},\ \bibinfo
  {pages} {104511} (\bibinfo {year} {2015}{\natexlab{a}})}\BibitemShut
  {NoStop}%
\bibitem [{\citenamefont {Wu}\ \emph {et~al.}(2015{\natexlab{b}})\citenamefont
  {Wu}, \citenamefont {Le}, \citenamefont {Yuan}, \citenamefont {Fan},\ and\
  \citenamefont {Hu}}]{wu.lee.15}%
  \BibitemOpen
  \bibfield  {author} {\bibinfo {author} {\bibfnamefont {X.-X.}\ \bibnamefont
  {Wu}}, \bibinfo {author} {\bibfnamefont {C.-C.}\ \bibnamefont {Le}}, \bibinfo
  {author} {\bibfnamefont {J.}~\bibnamefont {Yuan}}, \bibinfo {author}
  {\bibfnamefont {H.}~\bibnamefont {Fan}},\ and\ \bibinfo {author}
  {\bibfnamefont {J.-P.}\ \bibnamefont {Hu}},\ }\bibfield  {title} {\bibinfo
  {title} {Magnetism in quasi-one-dimensional {$A_{2}$Cr$_{3}$As$_{3}$}
  ({$A=$K, Rb}) superconductors},\ }\href
  {https://doi.org/10.1088/0256-307X/32/5/057401} {\bibfield  {journal}
  {\bibinfo  {journal} {Chinese Phys. Lett.}\ }\textbf {\bibinfo {volume}
  {32}},\ \bibinfo {pages} {057401} (\bibinfo {year}
  {2015}{\natexlab{b}})}\BibitemShut {NoStop}%
\bibitem [{\citenamefont {Jiang}\ \emph {et~al.}(2015)\citenamefont {Jiang},
  \citenamefont {Cao},\ and\ \citenamefont {Cao}}]{jiang.cao.15}%
  \BibitemOpen
  \bibfield  {author} {\bibinfo {author} {\bibfnamefont {H.}~\bibnamefont
  {Jiang}}, \bibinfo {author} {\bibfnamefont {G.}~\bibnamefont {Cao}},\ and\
  \bibinfo {author} {\bibfnamefont {C.}~\bibnamefont {Cao}},\ }\bibfield
  {title} {\bibinfo {title} {Electronic structure of quasi-one-dimensional
  superconductor {K$_{2}$Cr$_{3}$As$_{3}$} from first-principles
  calculations},\ }\href {https://doi.org/10.1038/srep16054} {\bibfield
  {journal} {\bibinfo  {journal} {Sci. Rep.}\ }\textbf {\bibinfo {volume}
  {5}},\ \bibinfo {pages} {16054} (\bibinfo {year} {2015})}\BibitemShut
  {NoStop}%
\bibitem [{\citenamefont {Xu}\ \emph {et~al.}(2020)\citenamefont {Xu},
  \citenamefont {Wu}, \citenamefont {Zhi}, \citenamefont {Lei}, \citenamefont
  {Duan}, \citenamefont {Ning}, \citenamefont {Cao},\ and\ \citenamefont
  {Chen}}]{xu.wu.20}%
  \BibitemOpen
  \bibfield  {author} {\bibinfo {author} {\bibfnamefont {C.}~\bibnamefont
  {Xu}}, \bibinfo {author} {\bibfnamefont {N.}~\bibnamefont {Wu}}, \bibinfo
  {author} {\bibfnamefont {G.-X.}\ \bibnamefont {Zhi}}, \bibinfo {author}
  {\bibfnamefont {B.-H.}\ \bibnamefont {Lei}}, \bibinfo {author} {\bibfnamefont
  {X.}~\bibnamefont {Duan}}, \bibinfo {author} {\bibfnamefont {F.}~\bibnamefont
  {Ning}}, \bibinfo {author} {\bibfnamefont {C.}~\bibnamefont {Cao}},\ and\
  \bibinfo {author} {\bibfnamefont {Q.}~\bibnamefont {Chen}},\ }\bibfield
  {title} {\bibinfo {title} {Coexistence of nontrivial topological properties
  and strong ferromagnetic fluctuations in quasi-one-dimensional
  {$A_{2}$Cr$_{3}$As$_{3}$}},\ }\href
  {https://doi.org/10.1038/s41524-020-0294-9} {\bibfield  {journal} {\bibinfo
  {journal} {npj Comput. Mater.}\ }\textbf {\bibinfo {volume} {6}},\ \bibinfo
  {pages} {30} (\bibinfo {year} {2020})}\BibitemShut {NoStop}%
\bibitem [{\citenamefont {Yang}\ \emph {et~al.}(2019)\citenamefont {Yang},
  \citenamefont {Feng}, \citenamefont {Lu}, \citenamefont {Wang},\ and\
  \citenamefont {Chen}}]{yang.feng.19}%
  \BibitemOpen
  \bibfield  {author} {\bibinfo {author} {\bibfnamefont {Y.}~\bibnamefont
  {Yang}}, \bibinfo {author} {\bibfnamefont {S.-Q.}\ \bibnamefont {Feng}},
  \bibinfo {author} {\bibfnamefont {H.-Y.}\ \bibnamefont {Lu}}, \bibinfo
  {author} {\bibfnamefont {W.-S.}\ \bibnamefont {Wang}},\ and\ \bibinfo
  {author} {\bibfnamefont {Z.-P.}\ \bibnamefont {Chen}},\ }\bibfield  {title}
  {\bibinfo {title} {Electronic structures of newly discovered
  quasi-one-dimensional superconductors {$A_{2}$Mo$_{3}$As$_{3}$} ({$A=$K, Rb,
  Cs})},\ }\href {https://doi.org/10.1007/s10948-019-5054-z} {\bibfield
  {journal} {\bibinfo  {journal} {J. Supercond. Nov. Magn.}\ }\textbf {\bibinfo
  {volume} {32}},\ \bibinfo {pages} {2421} (\bibinfo {year}
  {2019})}\BibitemShut {NoStop}%
\bibitem [{\citenamefont {Zhao}\ \emph {et~al.}(2020)\citenamefont {Zhao},
  \citenamefont {Mu}, \citenamefont {Ruan}, \citenamefont {Zhou}, \citenamefont
  {Yang}, \citenamefont {Liu}, \citenamefont {Pan}, \citenamefont {Zhang},
  \citenamefont {Chen},\ and\ \citenamefont {Ren}}]{zhao.mu.20}%
  \BibitemOpen
  \bibfield  {author} {\bibinfo {author} {\bibfnamefont {K.}~\bibnamefont
  {Zhao}}, \bibinfo {author} {\bibfnamefont {Q.-G.}\ \bibnamefont {Mu}},
  \bibinfo {author} {\bibfnamefont {B.-B.}\ \bibnamefont {Ruan}}, \bibinfo
  {author} {\bibfnamefont {M.-H.}\ \bibnamefont {Zhou}}, \bibinfo {author}
  {\bibfnamefont {Q.-S.}\ \bibnamefont {Yang}}, \bibinfo {author}
  {\bibfnamefont {T.}~\bibnamefont {Liu}}, \bibinfo {author} {\bibfnamefont
  {B.-J.}\ \bibnamefont {Pan}}, \bibinfo {author} {\bibfnamefont
  {S.}~\bibnamefont {Zhang}}, \bibinfo {author} {\bibfnamefont {G.-F.}\
  \bibnamefont {Chen}},\ and\ \bibinfo {author} {\bibfnamefont {Z.-A.}\
  \bibnamefont {Ren}},\ }\bibfield  {title} {\bibinfo {title} {A new
  quasi-one-dimensional ternary molybdenum pnictide {Rb$_{2}$Mo$_{3}$As$_{3}$}
  with superconducting transition at {10.5 K}},\ }\href
  {https://doi.org/10.1088/0256-307X/37/9/097401} {\bibfield  {journal}
  {\bibinfo  {journal} {Chinese Phys. Lett.}\ }\textbf {\bibinfo {volume}
  {37}},\ \bibinfo {pages} {097401} (\bibinfo {year} {2020})}\BibitemShut
  {NoStop}%
\bibitem [{\citenamefont {Lei}\ and\ \citenamefont
  {Singh}(2021)}]{lei.singh.21}%
  \BibitemOpen
  \bibfield  {author} {\bibinfo {author} {\bibfnamefont {B.-H.}\ \bibnamefont
  {Lei}}\ and\ \bibinfo {author} {\bibfnamefont {D.~J.}\ \bibnamefont
  {Singh}},\ }\bibfield  {title} {\bibinfo {title} {Multigap electron-phonon
  superconductivity in the quasi-one-dimensional pnictide
  {K$_{2}$Mo$_{3}$As$_{3}$}},\ }\href
  {https://doi.org/10.1103/PhysRevB.103.094512} {\bibfield  {journal} {\bibinfo
   {journal} {Phys. Rev. B}\ }\textbf {\bibinfo {volume} {103}},\ \bibinfo
  {pages} {094512} (\bibinfo {year} {2021})}\BibitemShut {NoStop}%
\bibitem [{\citenamefont {Barman}\ \emph {et~al.}(2020)\citenamefont {Barman},
  \citenamefont {Mondal}, \citenamefont {Pujari}, \citenamefont {Pathak},\ and\
  \citenamefont {Alam}}]{barman.mondal.20}%
  \BibitemOpen
  \bibfield  {author} {\bibinfo {author} {\bibfnamefont {C.~K.}\ \bibnamefont
  {Barman}}, \bibinfo {author} {\bibfnamefont {C.}~\bibnamefont {Mondal}},
  \bibinfo {author} {\bibfnamefont {S.}~\bibnamefont {Pujari}}, \bibinfo
  {author} {\bibfnamefont {B.}~\bibnamefont {Pathak}},\ and\ \bibinfo {author}
  {\bibfnamefont {A.}~\bibnamefont {Alam}},\ }\bibfield  {title} {\bibinfo
  {title} {Symmetry protection and giant {Fermi} arcs from multifold fermions
  in binary, ternary, and quaternary compounds},\ }\href
  {https://doi.org/10.1103/PhysRevB.102.155147} {\bibfield  {journal} {\bibinfo
   {journal} {Phys. Rev. B}\ }\textbf {\bibinfo {volume} {102}},\ \bibinfo
  {pages} {155147} (\bibinfo {year} {2020})}\BibitemShut {NoStop}%
\bibitem [{\citenamefont {Tsvyashchenko}\ \emph {et~al.}(2016)\citenamefont
  {Tsvyashchenko}, \citenamefont {Sidorov}, \citenamefont {Petrova},
  \citenamefont {Fomicheva}, \citenamefont {Zibrov},\ and\ \citenamefont
  {Dmitrienko}}]{tsvyashchenko.sidorov.16}%
  \BibitemOpen
  \bibfield  {author} {\bibinfo {author} {\bibfnamefont {A.}~\bibnamefont
  {Tsvyashchenko}}, \bibinfo {author} {\bibfnamefont {V.}~\bibnamefont
  {Sidorov}}, \bibinfo {author} {\bibfnamefont {A.}~\bibnamefont {Petrova}},
  \bibinfo {author} {\bibfnamefont {L.}~\bibnamefont {Fomicheva}}, \bibinfo
  {author} {\bibfnamefont {I.}~\bibnamefont {Zibrov}},\ and\ \bibinfo {author}
  {\bibfnamefont {V.}~\bibnamefont {Dmitrienko}},\ }\bibfield  {title}
  {\bibinfo {title} {Superconductivity and magnetism in noncentrosymmetric
  {RhGe}},\ }\href {https://doi.org/10.1016/j.jallcom.2016.06.048} {\bibfield
  {journal} {\bibinfo  {journal} {J. Alloys Compd.}\ }\textbf {\bibinfo
  {volume} {686}},\ \bibinfo {pages} {431} (\bibinfo {year}
  {2016})}\BibitemShut {NoStop}%
\bibitem [{\citenamefont {Fu}\ and\ \citenamefont {Ho}(1983)}]{fu.ho.83}%
  \BibitemOpen
  \bibfield  {author} {\bibinfo {author} {\bibfnamefont {C.~L.}\ \bibnamefont
  {Fu}}\ and\ \bibinfo {author} {\bibfnamefont {K.~M.}\ \bibnamefont {Ho}},\
  }\bibfield  {title} {\bibinfo {title} {First-principles calculation of the
  equilibrium ground-state properties of transition metals: Applications to
  {Nb} and {Mo}},\ }\href {https://doi.org/10.1103/PhysRevB.28.5480} {\bibfield
   {journal} {\bibinfo  {journal} {Phys. Rev. B}\ }\textbf {\bibinfo {volume}
  {28}},\ \bibinfo {pages} {5480} (\bibinfo {year} {1983})}\BibitemShut
  {NoStop}%
\bibitem [{\citenamefont {Ptok}\ \emph {et~al.}(2023)\citenamefont {Ptok},
  \citenamefont {Meier}, \citenamefont {Kobia\l{}ka}, \citenamefont {Basak},
  \citenamefont {Sternik}, \citenamefont {\L{}a\.{z}ewski}, \citenamefont
  {Jochym}, \citenamefont {McGuire}, \citenamefont {Sales}, \citenamefont
  {Miao}, \citenamefont {Piekarz},\ and\ \citenamefont
  {Ole\'{s}}}]{ptok.meier.23}%
  \BibitemOpen
  \bibfield  {author} {\bibinfo {author} {\bibfnamefont {A.}~\bibnamefont
  {Ptok}}, \bibinfo {author} {\bibfnamefont {W.~R.}\ \bibnamefont {Meier}},
  \bibinfo {author} {\bibfnamefont {A.}~\bibnamefont {Kobia\l{}ka}}, \bibinfo
  {author} {\bibfnamefont {S.}~\bibnamefont {Basak}}, \bibinfo {author}
  {\bibfnamefont {M.}~\bibnamefont {Sternik}}, \bibinfo {author} {\bibfnamefont
  {J.}~\bibnamefont {\L{}a\.{z}ewski}}, \bibinfo {author} {\bibfnamefont
  {P.~T.}\ \bibnamefont {Jochym}}, \bibinfo {author} {\bibfnamefont {M.~A.}\
  \bibnamefont {McGuire}}, \bibinfo {author} {\bibfnamefont {B.~C.}\
  \bibnamefont {Sales}}, \bibinfo {author} {\bibfnamefont {H.}~\bibnamefont
  {Miao}}, \bibinfo {author} {\bibfnamefont {P.}~\bibnamefont {Piekarz}},\ and\
  \bibinfo {author} {\bibfnamefont {A.~M.}\ \bibnamefont {Ole\'{s}}},\
  }\bibfield  {title} {\bibinfo {title} {Phononic drumhead surface state in the
  distorted kagome compound {RhPb}},\ }\href
  {https://doi.org/10.1103/PhysRevResearch.5.043231} {\bibfield  {journal}
  {\bibinfo  {journal} {Phys. Rev. Res.}\ }\textbf {\bibinfo {volume} {5}},\
  \bibinfo {pages} {043231} (\bibinfo {year} {2023})}\BibitemShut {NoStop}%
\bibitem [{\citenamefont {Tang}\ \emph {et~al.}(2017)\citenamefont {Tang},
  \citenamefont {Zhou},\ and\ \citenamefont {Zhang}}]{tang.zhou.17}%
  \BibitemOpen
  \bibfield  {author} {\bibinfo {author} {\bibfnamefont {P.}~\bibnamefont
  {Tang}}, \bibinfo {author} {\bibfnamefont {Q.}~\bibnamefont {Zhou}},\ and\
  \bibinfo {author} {\bibfnamefont {S.-C.}\ \bibnamefont {Zhang}},\ }\bibfield
  {title} {\bibinfo {title} {Multiple types of topological fermions in
  transition metal silicides},\ }\href
  {https://doi.org/10.1103/PhysRevLett.119.206402} {\bibfield  {journal}
  {\bibinfo  {journal} {Phys. Rev. Lett.}\ }\textbf {\bibinfo {volume} {119}},\
  \bibinfo {pages} {206402} (\bibinfo {year} {2017})}\BibitemShut {NoStop}%
\bibitem [{\citenamefont {Chang}\ \emph {et~al.}(2017)\citenamefont {Chang},
  \citenamefont {Xu}, \citenamefont {Wieder}, \citenamefont {Sanchez},
  \citenamefont {Huang}, \citenamefont {Belopolski}, \citenamefont {Chang},
  \citenamefont {Zhang}, \citenamefont {Bansil}, \citenamefont {Lin},\ and\
  \citenamefont {Hasan}}]{chang.xu.17}%
  \BibitemOpen
  \bibfield  {author} {\bibinfo {author} {\bibfnamefont {G.}~\bibnamefont
  {Chang}}, \bibinfo {author} {\bibfnamefont {S.-Y.}\ \bibnamefont {Xu}},
  \bibinfo {author} {\bibfnamefont {B.~J.}\ \bibnamefont {Wieder}}, \bibinfo
  {author} {\bibfnamefont {D.~S.}\ \bibnamefont {Sanchez}}, \bibinfo {author}
  {\bibfnamefont {S.-M.}\ \bibnamefont {Huang}}, \bibinfo {author}
  {\bibfnamefont {I.}~\bibnamefont {Belopolski}}, \bibinfo {author}
  {\bibfnamefont {T.-R.}\ \bibnamefont {Chang}}, \bibinfo {author}
  {\bibfnamefont {S.}~\bibnamefont {Zhang}}, \bibinfo {author} {\bibfnamefont
  {A.}~\bibnamefont {Bansil}}, \bibinfo {author} {\bibfnamefont
  {H.}~\bibnamefont {Lin}},\ and\ \bibinfo {author} {\bibfnamefont {M.~Z.}\
  \bibnamefont {Hasan}},\ }\bibfield  {title} {\bibinfo {title} {Unconventional
  chiral fermions and large topological {Fermi} arcs in {RhSi}},\ }\href
  {https://doi.org/10.1103/PhysRevLett.119.206401} {\bibfield  {journal}
  {\bibinfo  {journal} {Phys. Rev. Lett.}\ }\textbf {\bibinfo {volume} {119}},\
  \bibinfo {pages} {206401} (\bibinfo {year} {2017})}\BibitemShut {NoStop}%
\bibitem [{\citenamefont {Sanchez}\ \emph {et~al.}(2019)\citenamefont
  {Sanchez}, \citenamefont {Belopolski}, \citenamefont {Cochran}, \citenamefont
  {Xu}, \citenamefont {Yin}, \citenamefont {Chang}, \citenamefont {Xie},
  \citenamefont {Manna}, \citenamefont {S{\"u}{\ss}}, \citenamefont {Huang},
  \citenamefont {Alidoust}, \citenamefont {Multer}, \citenamefont {Zhang},
  \citenamefont {Shumiya}, \citenamefont {Wang}, \citenamefont {Wang},
  \citenamefont {Chang}, \citenamefont {Felser}, \citenamefont {Xu},
  \citenamefont {Jia}, \citenamefont {Lin},\ and\ \citenamefont
  {Hasan}}]{sanchez.balopolski.19}%
  \BibitemOpen
  \bibfield  {author} {\bibinfo {author} {\bibfnamefont {D.~S.}\ \bibnamefont
  {Sanchez}}, \bibinfo {author} {\bibfnamefont {I.}~\bibnamefont {Belopolski}},
  \bibinfo {author} {\bibfnamefont {T.~A.}\ \bibnamefont {Cochran}}, \bibinfo
  {author} {\bibfnamefont {X.}~\bibnamefont {Xu}}, \bibinfo {author}
  {\bibfnamefont {J.-X.}\ \bibnamefont {Yin}}, \bibinfo {author} {\bibfnamefont
  {G.}~\bibnamefont {Chang}}, \bibinfo {author} {\bibfnamefont
  {W.}~\bibnamefont {Xie}}, \bibinfo {author} {\bibfnamefont {K.}~\bibnamefont
  {Manna}}, \bibinfo {author} {\bibfnamefont {V.}~\bibnamefont {S{\"u}{\ss}}},
  \bibinfo {author} {\bibfnamefont {C.-Y.}\ \bibnamefont {Huang}}, \bibinfo
  {author} {\bibfnamefont {N.}~\bibnamefont {Alidoust}}, \bibinfo {author}
  {\bibfnamefont {D.}~\bibnamefont {Multer}}, \bibinfo {author} {\bibfnamefont
  {S.~S.}\ \bibnamefont {Zhang}}, \bibinfo {author} {\bibfnamefont
  {N.}~\bibnamefont {Shumiya}}, \bibinfo {author} {\bibfnamefont
  {X.}~\bibnamefont {Wang}}, \bibinfo {author} {\bibfnamefont {G.-Q.}\
  \bibnamefont {Wang}}, \bibinfo {author} {\bibfnamefont {T.-R.}\ \bibnamefont
  {Chang}}, \bibinfo {author} {\bibfnamefont {C.}~\bibnamefont {Felser}},
  \bibinfo {author} {\bibfnamefont {S.-Y.}\ \bibnamefont {Xu}}, \bibinfo
  {author} {\bibfnamefont {S.}~\bibnamefont {Jia}}, \bibinfo {author}
  {\bibfnamefont {H.}~\bibnamefont {Lin}},\ and\ \bibinfo {author}
  {\bibfnamefont {M.~Z.}\ \bibnamefont {Hasan}},\ }\bibfield  {title} {\bibinfo
  {title} {Topological chiral crystals with helicoid-arc quantum states},\
  }\href {https://doi.org/10.1038/s41586-019-1037-2} {\bibfield  {journal}
  {\bibinfo  {journal} {Nature}\ }\textbf {\bibinfo {volume} {567}},\ \bibinfo
  {pages} {500} (\bibinfo {year} {2019})}\BibitemShut {NoStop}%
\bibitem [{\citenamefont {Takane}\ \emph {et~al.}(2019)\citenamefont {Takane},
  \citenamefont {Wang}, \citenamefont {Souma}, \citenamefont {Nakayama},
  \citenamefont {Nakamura}, \citenamefont {Oinuma}, \citenamefont {Nakata},
  \citenamefont {Iwasawa}, \citenamefont {Cacho}, \citenamefont {Kim},
  \citenamefont {Horiba}, \citenamefont {Kumigashira}, \citenamefont
  {Takahashi}, \citenamefont {Ando},\ and\ \citenamefont
  {Sato}}]{takane.wang.19}%
  \BibitemOpen
  \bibfield  {author} {\bibinfo {author} {\bibfnamefont {D.}~\bibnamefont
  {Takane}}, \bibinfo {author} {\bibfnamefont {Z.}~\bibnamefont {Wang}},
  \bibinfo {author} {\bibfnamefont {S.}~\bibnamefont {Souma}}, \bibinfo
  {author} {\bibfnamefont {K.}~\bibnamefont {Nakayama}}, \bibinfo {author}
  {\bibfnamefont {T.}~\bibnamefont {Nakamura}}, \bibinfo {author}
  {\bibfnamefont {H.}~\bibnamefont {Oinuma}}, \bibinfo {author} {\bibfnamefont
  {Y.}~\bibnamefont {Nakata}}, \bibinfo {author} {\bibfnamefont
  {H.}~\bibnamefont {Iwasawa}}, \bibinfo {author} {\bibfnamefont
  {C.}~\bibnamefont {Cacho}}, \bibinfo {author} {\bibfnamefont
  {T.}~\bibnamefont {Kim}}, \bibinfo {author} {\bibfnamefont {K.}~\bibnamefont
  {Horiba}}, \bibinfo {author} {\bibfnamefont {H.}~\bibnamefont {Kumigashira}},
  \bibinfo {author} {\bibfnamefont {T.}~\bibnamefont {Takahashi}}, \bibinfo
  {author} {\bibfnamefont {Y.}~\bibnamefont {Ando}},\ and\ \bibinfo {author}
  {\bibfnamefont {T.}~\bibnamefont {Sato}},\ }\bibfield  {title} {\bibinfo
  {title} {Observation of chiral fermions with a large topological charge and
  associated {Fermi}-arc surface states in {CoSi}},\ }\href
  {https://doi.org/10.1103/PhysRevLett.122.076402} {\bibfield  {journal}
  {\bibinfo  {journal} {Phys. Rev. Lett.}\ }\textbf {\bibinfo {volume} {122}},\
  \bibinfo {pages} {076402} (\bibinfo {year} {2019})}\BibitemShut {NoStop}%
\bibitem [{\citenamefont {Schr{\"o}ter}\ \emph {et~al.}(2019)\citenamefont
  {Schr{\"o}ter}, \citenamefont {Pei}, \citenamefont {Vergniory}, \citenamefont
  {Sun}, \citenamefont {Manna}, \citenamefont {de~Juan}, \citenamefont
  {Krieger}, \citenamefont {S{\"u}ss}, \citenamefont {Schmidt}, \citenamefont
  {Dudin}, \citenamefont {Bradlyn}, \citenamefont {Kim}, \citenamefont
  {Schmitt}, \citenamefont {Cacho}, \citenamefont {Felser}, \citenamefont
  {Strocov},\ and\ \citenamefont {Chen}}]{schroter.pei.19}%
  \BibitemOpen
  \bibfield  {author} {\bibinfo {author} {\bibfnamefont {N.~B.~M.}\
  \bibnamefont {Schr{\"o}ter}}, \bibinfo {author} {\bibfnamefont
  {D.}~\bibnamefont {Pei}}, \bibinfo {author} {\bibfnamefont {M.~G.}\
  \bibnamefont {Vergniory}}, \bibinfo {author} {\bibfnamefont {Y.}~\bibnamefont
  {Sun}}, \bibinfo {author} {\bibfnamefont {K.}~\bibnamefont {Manna}}, \bibinfo
  {author} {\bibfnamefont {F.}~\bibnamefont {de~Juan}}, \bibinfo {author}
  {\bibfnamefont {J.~A.}\ \bibnamefont {Krieger}}, \bibinfo {author}
  {\bibfnamefont {V.}~\bibnamefont {S{\"u}ss}}, \bibinfo {author}
  {\bibfnamefont {M.}~\bibnamefont {Schmidt}}, \bibinfo {author} {\bibfnamefont
  {P.}~\bibnamefont {Dudin}}, \bibinfo {author} {\bibfnamefont
  {B.}~\bibnamefont {Bradlyn}}, \bibinfo {author} {\bibfnamefont {T.~K.}\
  \bibnamefont {Kim}}, \bibinfo {author} {\bibfnamefont {T.}~\bibnamefont
  {Schmitt}}, \bibinfo {author} {\bibfnamefont {C.}~\bibnamefont {Cacho}},
  \bibinfo {author} {\bibfnamefont {C.}~\bibnamefont {Felser}}, \bibinfo
  {author} {\bibfnamefont {V.~N.}\ \bibnamefont {Strocov}},\ and\ \bibinfo
  {author} {\bibfnamefont {Y.}~\bibnamefont {Chen}},\ }\bibfield  {title}
  {\bibinfo {title} {Chiral topological semimetal with multifold band crossings
  and long {Fermi} arcs},\ }\href {https://doi.org/10.1038/s41567-019-0511-y}
  {\bibfield  {journal} {\bibinfo  {journal} {Nat. Phys.}\ }\textbf {\bibinfo
  {volume} {15}},\ \bibinfo {pages} {759} (\bibinfo {year} {2019})}\BibitemShut
  {NoStop}%
\bibitem [{\citenamefont {Li}\ \emph {et~al.}(2019)\citenamefont {Li},
  \citenamefont {Xu}, \citenamefont {Rao}, \citenamefont {Zhou}, \citenamefont
  {Wang}, \citenamefont {Zhou}, \citenamefont {Tian}, \citenamefont {Gao},
  \citenamefont {Li}, \citenamefont {Huang}, \citenamefont {Lei}, \citenamefont
  {Weng}, \citenamefont {Sun}, \citenamefont {Xia}, \citenamefont {Qian},\ and\
  \citenamefont {Ding}}]{li.xu.19}%
  \BibitemOpen
  \bibfield  {author} {\bibinfo {author} {\bibfnamefont {H.}~\bibnamefont
  {Li}}, \bibinfo {author} {\bibfnamefont {S.}~\bibnamefont {Xu}}, \bibinfo
  {author} {\bibfnamefont {Z.-C.}\ \bibnamefont {Rao}}, \bibinfo {author}
  {\bibfnamefont {L.-Q.}\ \bibnamefont {Zhou}}, \bibinfo {author}
  {\bibfnamefont {Z.-J.}\ \bibnamefont {Wang}}, \bibinfo {author}
  {\bibfnamefont {S.-M.}\ \bibnamefont {Zhou}}, \bibinfo {author}
  {\bibfnamefont {S.-J.}\ \bibnamefont {Tian}}, \bibinfo {author}
  {\bibfnamefont {S.-Y.}\ \bibnamefont {Gao}}, \bibinfo {author} {\bibfnamefont
  {J.-J.}\ \bibnamefont {Li}}, \bibinfo {author} {\bibfnamefont {Y.-B.}\
  \bibnamefont {Huang}}, \bibinfo {author} {\bibfnamefont {H.-C.}\ \bibnamefont
  {Lei}}, \bibinfo {author} {\bibfnamefont {H.-M.}\ \bibnamefont {Weng}},
  \bibinfo {author} {\bibfnamefont {Y.-J.}\ \bibnamefont {Sun}}, \bibinfo
  {author} {\bibfnamefont {T.-L.}\ \bibnamefont {Xia}}, \bibinfo {author}
  {\bibfnamefont {T.}~\bibnamefont {Qian}},\ and\ \bibinfo {author}
  {\bibfnamefont {H.}~\bibnamefont {Ding}},\ }\bibfield  {title} {\bibinfo
  {title} {Chiral fermion reversal in chiral crystals},\ }\href
  {https://doi.org/10.1038/s41467-019-13435-4} {\bibfield  {journal} {\bibinfo
  {journal} {Nat. Commun.}\ }\textbf {\bibinfo {volume} {10}},\ \bibinfo
  {pages} {5505} (\bibinfo {year} {2019})}\BibitemShut {NoStop}%
\bibitem [{\citenamefont {Yao}\ \emph {et~al.}(2020)\citenamefont {Yao},
  \citenamefont {Manna}, \citenamefont {Yang}, \citenamefont {Fedorov},
  \citenamefont {Voroshnin}, \citenamefont {Valentin~Schwarze}, \citenamefont
  {Hornung}, \citenamefont {Chattopadhyay}, \citenamefont {Sun}, \citenamefont
  {Guin}, \citenamefont {Wosnitza}, \citenamefont {Borrmann}, \citenamefont
  {Shekhar}, \citenamefont {Kumar}, \citenamefont {Fink}, \citenamefont {Sun},\
  and\ \citenamefont {Felser}}]{yao.manna.20}%
  \BibitemOpen
  \bibfield  {author} {\bibinfo {author} {\bibfnamefont {M.}~\bibnamefont
  {Yao}}, \bibinfo {author} {\bibfnamefont {K.}~\bibnamefont {Manna}}, \bibinfo
  {author} {\bibfnamefont {Q.}~\bibnamefont {Yang}}, \bibinfo {author}
  {\bibfnamefont {A.}~\bibnamefont {Fedorov}}, \bibinfo {author} {\bibfnamefont
  {V.}~\bibnamefont {Voroshnin}}, \bibinfo {author} {\bibfnamefont
  {B.}~\bibnamefont {Valentin~Schwarze}}, \bibinfo {author} {\bibfnamefont
  {J.}~\bibnamefont {Hornung}}, \bibinfo {author} {\bibfnamefont
  {S.}~\bibnamefont {Chattopadhyay}}, \bibinfo {author} {\bibfnamefont
  {Z.}~\bibnamefont {Sun}}, \bibinfo {author} {\bibfnamefont {S.~N.}\
  \bibnamefont {Guin}}, \bibinfo {author} {\bibfnamefont {J.}~\bibnamefont
  {Wosnitza}}, \bibinfo {author} {\bibfnamefont {H.}~\bibnamefont {Borrmann}},
  \bibinfo {author} {\bibfnamefont {C.}~\bibnamefont {Shekhar}}, \bibinfo
  {author} {\bibfnamefont {N.}~\bibnamefont {Kumar}}, \bibinfo {author}
  {\bibfnamefont {J.}~\bibnamefont {Fink}}, \bibinfo {author} {\bibfnamefont
  {Y.}~\bibnamefont {Sun}},\ and\ \bibinfo {author} {\bibfnamefont
  {C.}~\bibnamefont {Felser}},\ }\bibfield  {title} {\bibinfo {title}
  {Observation of giant spin-split {Fermi}-arc with maximal chern number in the
  chiral topological semimetal {PtGa}},\ }\href
  {https://doi.org/10.1038/s41467-020-15865-x} {\bibfield  {journal} {\bibinfo
  {journal} {Nat. Commun.}\ }\textbf {\bibinfo {volume} {11}},\ \bibinfo
  {pages} {2033} (\bibinfo {year} {2020})}\BibitemShut {NoStop}%
\bibitem [{\citenamefont {Bose}\ and\ \citenamefont
  {Narayan}(2021)}]{bose.narayan.21}%
  \BibitemOpen
  \bibfield  {author} {\bibinfo {author} {\bibfnamefont {A.}~\bibnamefont
  {Bose}}\ and\ \bibinfo {author} {\bibfnamefont {A.}~\bibnamefont {Narayan}},\
  }\bibfield  {title} {\bibinfo {title} {Strain-induced topological charge
  control in multifold fermion systems},\ }\href
  {https://doi.org/10.1088/1361-648X/ac0fa0} {\bibfield  {journal} {\bibinfo
  {journal} {J. Phys.: Condens. Matter}\ }\textbf {\bibinfo {volume} {33}},\
  \bibinfo {pages} {375002} (\bibinfo {year} {2021})}\BibitemShut {NoStop}%
\bibitem [{\citenamefont {Lifshitz}(1960)}]{lifshitz.60}%
  \BibitemOpen
  \bibfield  {author} {\bibinfo {author} {\bibfnamefont {I.~M.}\ \bibnamefont
  {Lifshitz}},\ }\bibfield  {title} {\bibinfo {title} {Anomalies of electron
  characteristics of a metal in the high pressure region},\ }\href@noop {}
  {\bibfield  {journal} {\bibinfo  {journal} {Zh. Eksp. Teor. Fiz.}\ }\textbf
  {\bibinfo {volume} {38}},\ \bibinfo {pages} {1569} (\bibinfo {year}
  {1960})},\ \bibinfo {note} {[Sov. Phys. JETP \textbf{11}, 1130--1135
  (1960)]}\BibitemShut {NoStop}%
\bibitem [{\citenamefont {Momma}\ and\ \citenamefont
  {Izumi}(2011)}]{momma.izumi.11}%
  \BibitemOpen
  \bibfield  {author} {\bibinfo {author} {\bibfnamefont {K.}~\bibnamefont
  {Momma}}\ and\ \bibinfo {author} {\bibfnamefont {F.}~\bibnamefont {Izumi}},\
  }\bibfield  {title} {\bibinfo {title} {{{\sc vesta3} for three-dimensional
  visualization of crystal, volumetric and morphology data}},\ }\href
  {https://doi.org/10.1107/S0021889811038970} {\bibfield  {journal} {\bibinfo
  {journal} {J. Appl. Crystallogr.}\ }\textbf {\bibinfo {volume} {44}},\
  \bibinfo {pages} {1272} (\bibinfo {year} {2011})}\BibitemShut {NoStop}%
\bibitem [{\citenamefont {Kokalj}(1999)}]{kokalj.99}%
  \BibitemOpen
  \bibfield  {author} {\bibinfo {author} {\bibfnamefont {A.}~\bibnamefont
  {Kokalj}},\ }\bibfield  {title} {\bibinfo {title} {Xcrysden--a new program
  for displaying crystalline structures and electron densities},\ }\href
  {https://doi.org/10.1016/S1093-3263(99)00028-5} {\bibfield  {journal}
  {\bibinfo  {journal} {J. Mol. Graph. Model.}\ }\textbf {\bibinfo {volume}
  {17}},\ \bibinfo {pages} {176} (\bibinfo {year} {1999})}\BibitemShut
  {NoStop}%
\bibitem [{\citenamefont {Kresse}\ and\ \citenamefont
  {Hafner}(1994)}]{kresse.hafner.94}%
  \BibitemOpen
  \bibfield  {author} {\bibinfo {author} {\bibfnamefont {G.}~\bibnamefont
  {Kresse}}\ and\ \bibinfo {author} {\bibfnamefont {J.}~\bibnamefont
  {Hafner}},\ }\bibfield  {title} {\bibinfo {title} {Ab initio
  molecular-dynamics simulation of the liquid-metal--amorphous-semiconductor
  transition in germanium},\ }\href {https://doi.org/10.1103/PhysRevB.49.14251}
  {\bibfield  {journal} {\bibinfo  {journal} {Phys. Rev. B}\ }\textbf {\bibinfo
  {volume} {49}},\ \bibinfo {pages} {14251} (\bibinfo {year}
  {1994})}\BibitemShut {NoStop}%
\bibitem [{\citenamefont {Kresse}\ and\ \citenamefont
  {Furthm\"uller}(1996)}]{kresse.furthmuller.96}%
  \BibitemOpen
  \bibfield  {author} {\bibinfo {author} {\bibfnamefont {G.}~\bibnamefont
  {Kresse}}\ and\ \bibinfo {author} {\bibfnamefont {J.}~\bibnamefont
  {Furthm\"uller}},\ }\bibfield  {title} {\bibinfo {title} {Efficient iterative
  schemes for ab initio total-energy calculations using a plane-wave basis
  set},\ }\href {https://doi.org/10.1103/PhysRevB.54.11169} {\bibfield
  {journal} {\bibinfo  {journal} {Phys. Rev. B}\ }\textbf {\bibinfo {volume}
  {54}},\ \bibinfo {pages} {11169} (\bibinfo {year} {1996})}\BibitemShut
  {NoStop}%
\bibitem [{\citenamefont {Kresse}\ and\ \citenamefont
  {Joubert}(1999)}]{kresse.joubert.99}%
  \BibitemOpen
  \bibfield  {author} {\bibinfo {author} {\bibfnamefont {G.}~\bibnamefont
  {Kresse}}\ and\ \bibinfo {author} {\bibfnamefont {D.}~\bibnamefont
  {Joubert}},\ }\bibfield  {title} {\bibinfo {title} {From ultrasoft
  pseudopotentials to the projector augmented-wave method},\ }\href
  {https://doi.org/10.1103/PhysRevB.59.1758} {\bibfield  {journal} {\bibinfo
  {journal} {Phys. Rev. B}\ }\textbf {\bibinfo {volume} {59}},\ \bibinfo
  {pages} {1758} (\bibinfo {year} {1999})}\BibitemShut {NoStop}%
\bibitem [{\citenamefont {Bl\"ochl}(1994)}]{blochl.94}%
  \BibitemOpen
  \bibfield  {author} {\bibinfo {author} {\bibfnamefont {P.~E.}\ \bibnamefont
  {Bl\"ochl}},\ }\bibfield  {title} {\bibinfo {title} {Projector augmented-wave
  method},\ }\href {https://doi.org/10.1103/PhysRevB.50.17953} {\bibfield
  {journal} {\bibinfo  {journal} {Phys. Rev. B}\ }\textbf {\bibinfo {volume}
  {50}},\ \bibinfo {pages} {17953} (\bibinfo {year} {1994})}\BibitemShut
  {NoStop}%
\bibitem [{\citenamefont {Perdew}\ \emph {et~al.}(2008)\citenamefont {Perdew},
  \citenamefont {Ruzsinszky}, \citenamefont {Csonka}, \citenamefont {Vydrov},
  \citenamefont {Scuseria}, \citenamefont {Constantin}, \citenamefont {Zhou},\
  and\ \citenamefont {Burke}}]{perdew.ruzsinszky.08}%
  \BibitemOpen
  \bibfield  {author} {\bibinfo {author} {\bibfnamefont {J.~P.}\ \bibnamefont
  {Perdew}}, \bibinfo {author} {\bibfnamefont {A.}~\bibnamefont {Ruzsinszky}},
  \bibinfo {author} {\bibfnamefont {G.~I.}\ \bibnamefont {Csonka}}, \bibinfo
  {author} {\bibfnamefont {O.~A.}\ \bibnamefont {Vydrov}}, \bibinfo {author}
  {\bibfnamefont {G.~E.}\ \bibnamefont {Scuseria}}, \bibinfo {author}
  {\bibfnamefont {L.~A.}\ \bibnamefont {Constantin}}, \bibinfo {author}
  {\bibfnamefont {X.}~\bibnamefont {Zhou}},\ and\ \bibinfo {author}
  {\bibfnamefont {K.}~\bibnamefont {Burke}},\ }\bibfield  {title} {\bibinfo
  {title} {Restoring the density-gradient expansion for exchange in solids and
  surfaces},\ }\href {https://doi.org/10.1103/PhysRevLett.100.136406}
  {\bibfield  {journal} {\bibinfo  {journal} {Phys. Rev. Lett.}\ }\textbf
  {\bibinfo {volume} {100}},\ \bibinfo {pages} {136406} (\bibinfo {year}
  {2008})}\BibitemShut {NoStop}%
\bibitem [{\citenamefont {Monkhorst}\ and\ \citenamefont
  {Pack}(1976)}]{monkhorst.pack.76}%
  \BibitemOpen
  \bibfield  {author} {\bibinfo {author} {\bibfnamefont {H.~J.}\ \bibnamefont
  {Monkhorst}}\ and\ \bibinfo {author} {\bibfnamefont {J.~D.}\ \bibnamefont
  {Pack}},\ }\bibfield  {title} {\bibinfo {title} {Special points for
  {Brillouin}-zone integrations},\ }\href
  {https://doi.org/10.1103/PhysRevB.13.5188} {\bibfield  {journal} {\bibinfo
  {journal} {Phys. Rev. B}\ }\textbf {\bibinfo {volume} {13}},\ \bibinfo
  {pages} {5188} (\bibinfo {year} {1976})}\BibitemShut {NoStop}%
\bibitem [{\citenamefont {Stokes}\ and\ \citenamefont
  {Hatch}(2005)}]{stokes.hatch.05}%
  \BibitemOpen
  \bibfield  {author} {\bibinfo {author} {\bibfnamefont {H.~T.}\ \bibnamefont
  {Stokes}}\ and\ \bibinfo {author} {\bibfnamefont {D.~M.}\ \bibnamefont
  {Hatch}},\ }\bibfield  {title} {\bibinfo {title} {{{\sc FindSym}: program for
  identifying the space-group symmetry of a crystal}},\ }\href
  {https://doi.org/10.1107/S0021889804031528} {\bibfield  {journal} {\bibinfo
  {journal} {J. Appl. Cryst.}\ }\textbf {\bibinfo {volume} {38}},\ \bibinfo
  {pages} {237} (\bibinfo {year} {2005})}\BibitemShut {NoStop}%
\bibitem [{\citenamefont {Togo}\ and\ \citenamefont
  {Tanaka}(2018)}]{togo.tanaka.18}%
  \BibitemOpen
  \bibfield  {author} {\bibinfo {author} {\bibfnamefont {A.}~\bibnamefont
  {Togo}}\ and\ \bibinfo {author} {\bibfnamefont {I.}~\bibnamefont {Tanaka}},\
  }\href@noop {} {\bibinfo {title} {{\sc Spglib}: a software library for
  crystal symmetry search}} (\bibinfo {year} {2018}),\ \Eprint
  {https://arxiv.org/abs/arXiv:1808.01590} {arXiv:1808.01590} \BibitemShut
  {NoStop}%
\bibitem [{\citenamefont {Hinuma}\ \emph {et~al.}(2017)\citenamefont {Hinuma},
  \citenamefont {Pizzi}, \citenamefont {Kumagai}, \citenamefont {Oba},\ and\
  \citenamefont {Tanaka}}]{hinuma.pizzi.17}%
  \BibitemOpen
  \bibfield  {author} {\bibinfo {author} {\bibfnamefont {Y.}~\bibnamefont
  {Hinuma}}, \bibinfo {author} {\bibfnamefont {G.}~\bibnamefont {Pizzi}},
  \bibinfo {author} {\bibfnamefont {Y.}~\bibnamefont {Kumagai}}, \bibinfo
  {author} {\bibfnamefont {F.}~\bibnamefont {Oba}},\ and\ \bibinfo {author}
  {\bibfnamefont {I.}~\bibnamefont {Tanaka}},\ }\bibfield  {title} {\bibinfo
  {title} {Band structure diagram paths based on crystallography},\ }\href
  {https://doi.org/10.1016/j.commatsci.2016.10.015} {\bibfield  {journal}
  {\bibinfo  {journal} {Comput. Mater. Sci.}\ }\textbf {\bibinfo {volume}
  {128}},\ \bibinfo {pages} {140} (\bibinfo {year} {2017})}\BibitemShut
  {NoStop}%
\bibitem [{\citenamefont {Parlinski}\ \emph {et~al.}(1997)\citenamefont
  {Parlinski}, \citenamefont {Li},\ and\ \citenamefont
  {Kawazoe}}]{parlinski.li.97}%
  \BibitemOpen
  \bibfield  {author} {\bibinfo {author} {\bibfnamefont {K.}~\bibnamefont
  {Parlinski}}, \bibinfo {author} {\bibfnamefont {Z.~Q.}\ \bibnamefont {Li}},\
  and\ \bibinfo {author} {\bibfnamefont {Y.}~\bibnamefont {Kawazoe}},\
  }\bibfield  {title} {\bibinfo {title} {First-principles determination of the
  soft mode in cubic {ZrO$_{2}$}},\ }\href
  {https://doi.org/10.1103/PhysRevLett.78.4063} {\bibfield  {journal} {\bibinfo
   {journal} {Phys. Rev. Lett.}\ }\textbf {\bibinfo {volume} {78}},\ \bibinfo
  {pages} {4063} (\bibinfo {year} {1997})}\BibitemShut {NoStop}%
\bibitem [{\citenamefont {Togo}\ \emph {et~al.}(2023)\citenamefont {Togo},
  \citenamefont {Chaput}, \citenamefont {Tadano},\ and\ \citenamefont
  {Tanaka}}]{togo.chaput.23}%
  \BibitemOpen
  \bibfield  {author} {\bibinfo {author} {\bibfnamefont {A.}~\bibnamefont
  {Togo}}, \bibinfo {author} {\bibfnamefont {L.}~\bibnamefont {Chaput}},
  \bibinfo {author} {\bibfnamefont {T.}~\bibnamefont {Tadano}},\ and\ \bibinfo
  {author} {\bibfnamefont {I.}~\bibnamefont {Tanaka}},\ }\bibfield  {title}
  {\bibinfo {title} {Implementation strategies in phonopy and phono3py},\
  }\href {https://doi.org/10.1088/1361-648X/acd831} {\bibfield  {journal}
  {\bibinfo  {journal} {J. Phys. Condens. Matter}\ }\textbf {\bibinfo {volume}
  {35}},\ \bibinfo {pages} {353001} (\bibinfo {year} {2023})}\BibitemShut
  {NoStop}%
\bibitem [{\citenamefont {Togo}(2023)}]{togo.23}%
  \BibitemOpen
  \bibfield  {author} {\bibinfo {author} {\bibfnamefont {A.}~\bibnamefont
  {Togo}},\ }\bibfield  {title} {\bibinfo {title} {First-principles phonon
  calculations with phonopy and phono3py},\ }\href
  {https://doi.org/10.7566/JPSJ.92.012001} {\bibfield  {journal} {\bibinfo
  {journal} {J. Phys. Soc. Jpn.}\ }\textbf {\bibinfo {volume} {92}},\ \bibinfo
  {pages} {012001} (\bibinfo {year} {2023})}\BibitemShut {NoStop}%
\end{thebibliography}%

\end{document}